\documentclass[prd, aps, showpacs]{revtex4}

\usepackage{latexsym, graphicx}

\begin{document}

\title{Accelerated Detector - Quantum Field Correlations: \\
From Vacuum Fluctuations to Radiation Flux}
\author{Shih-Yuin Lin}
\email{sylin@phys.sinica.edu.tw}
\author{B. L. Hu}
\email{hub@physics.umd.edu}
\affiliation{Center for Quantum and Gravitational Physics,
Institute of Physics, Academia Sinica, Nankang, Taipei 11529, Taiwan\\
and Department of Physics, University of Maryland, College Park,
Maryland 20742-4111, USA}
\date{April 3, 2006}

\begin{abstract}
In this paper we analyze the interaction of a uniformly accelerated
detector with a quantum field in (3+1)D spacetime, aiming at
the issue of how kinematics can render vacuum fluctuations the
appearance of thermal radiance in the detector (Unruh effect) and
how they engender flux of radiation for observers afar.
Two basic questions are addressed in this study:  a) How are
vacuum fluctuations related to the emitted radiation? b) Is there
emitted radiation with energy flux in the Unruh effect? 
We adopt a method which places the detector and the field on an
equal footing and derive the two-point correlation functions of the
detector and of the field separately with full account of their
interplay. From the exact solutions, we are able to study the
complete process from the initial transient to the final steady
state, keeping track of all activities they engage in and the
physical effects manifested.
We derive a quantum radiation formula for a Minkowski observer. We
find that there does exist a positive radiated power of quantum
nature emitted by the detector, with a hint of certain features of
the Unruh effect. We further verify that the total energy of the
dressed detector and a part of the radiated energy from
the detector is conserved. However, this part of the radiation ceases
in steady state. So the hint of the Unruh effect in radiated power
is actually not directly from the energy flux that the detector
experiences in Unruh effect.
Since all the relevant
quantum and statistical information about the detector (atom) and
the field can be obtained from the results presented here, they
are expected to be useful, when appropriately generalized, for
addressing issues of quantum information processing in atomic and
optical systems, such as quantum decoherence, entanglement and
teleportation.
\end{abstract}


\maketitle

\section{Introduction}

Inasmuch as studies of the interaction between a particle and a
quantum field are basic to particle physics and field theory,
understanding the interaction between an atom and a quantum field is
essential to current atomic and optical physics research
\cite{CTDG,CTmovatom,Milonni,CPP,scully}. The interaction of an
accelerated charge or detector (an object with some internal degrees
of freedom such as an atom or harmonic oscillator) in a quantum
field is a simple yet fundamental problem with many implications in
quantum field theory \cite{BD,DeW75}, thermodynamics \cite{BH, Haw75}
and applications in radiation theory and atomic-optical physics.

It is common knowledge that accelerating charges give rise to
radiation \cite{Jackson}. But it is not entirely straightforward to 
derive the radiation formula from quantum field theory. 
How are vacuum fluctuations related to the emitted radiation?
When an atom or detector moves at constant 
acceleration, according to Unruh \cite{Unr76}, it would experience 
a thermal bath at temperature $T_U= \hbar a /(2\pi c k_B)$, where 
$a$ is the proper acceleration. Is there emitted radiation
with an energy flux in the Unruh effect?

Unruh effect, and the related effect for moving mirrors studied by
Davies and Fulling \cite{DavFul}, were intended originally to mimic
Hawking radiation from black holes. Because of this connection, for
some time now there has been a speculation 
that there is real radiation emitted from a uniformly accelerated
detector (UAD) under steady state conditions (i.e., for atoms
which have been uniformly accelerated for a time sufficiently long
that transient effects have died out), not unlike that of an
accelerating charge \cite{Jackson, Boulware}.
In light of pending experiments both for electrons in accelerators
\cite{Chen,ChenTaj,BelLen} and for accelerated atoms in optical
cavities \cite{Scullyetal} this speculation 
has acquired some realistic significance. There is a need for more
detailed analysis for both the uniform acceleration of charges or
detectors and for transient motions because the latter can produce
radiation and as explained below, sudden changes in the dynamics can
also produce emitted radiation with thermal characteristics.

After Unruh and Wald's \cite{UnrWal} earlier explication of what a
Minkowski observer sees, Grove \cite{Grove} questioned whether an
accelerated atom actually emits radiated energy. Raine, Sciama and
Grove \cite{RSG} (RSG) analyzed what an inertial observer placed in
the forward light cone of the accelerating detector would measure
and concluded that the oscillator does not radiate. Unruh
\cite{Unr92}, in an independent calculation, basically concurred
with the findings of RSG but he also showed the existence of extra
terms in the two-point function of the field which would contribute
to the excitation of a detector placed in the forward light cone.
Massar, Parantani and Brout \cite{MPB93} (MPB) pointed out that the
missing terms in RSG contribute to a ``polarization cloud" around the
accelerating detector. For a review of earlier work on accelerated
detectors, see e.g., \cite{AccDetRev}. For work after that, see,
e.g., Hinterleitner \cite{Hin}, Audretsch, M\"{u}ller and Holzmann
\cite{AMH}, Massar and Parantani \cite{MP}. Our present work
follows the vein of Raval, Hu, Anglin (RHA) and Koks
\cite{RavalPhD,RHA,RHK,CapHR} on the minimal coupling model and uses
some results of Lin \cite{lin03b} on the Unruh-DeWitt model
\cite{Unr76,DeW79}.

With regard to the question ``Is there a radiation flux emitted from
an Unruh detector?" the findings of RSG, Unruh, MPB, RHA and others
show that, at least in (1+1) dimension model calculations,
\textit{there is no emitted radiation from a linear uniformly
accelerated oscillator under equilibrium conditions}, even though,
as found before, that there exists a polarization cloud around it.
Hu and Johnson \cite{CapHJ} emphasized the difference between an
equilibrium condition (steady state uniform acceleration) where
there is no emitted radiation, and nonequilibrium conditions where
there could be radiation emitted. Nonequilibrium conditions arise
for non-uniformly accelerated atoms (for an example of finite time
acceleration, see Raval, Hu and Koks (RHK) \cite{RHK}), or during
the initial transient time for an atom approaching uniform
acceleration, when its internal states have not yet reached
equilibrium through interaction with the field. Hu and Raval (HR)
\cite{CapHR,RavalPhD} presented a more complete analysis of the
two-point function, calculated for two points lying in arbitrary
regions of Minkowski space. This generalizes the results of MPB in
that there is no restriction for the two points to lie to the left
of the accelerated oscillator trajectory. They show where the extra
terms in the two-point function are which were ignored in the RSG
analysis. More important to answering the theme question, they show
that at least in (1+1) dimension the stress-energy tensor vanishes
everywhere except on the horizon. This means that there is no net
flux of radiation emitted from the uniformly accelerated oscillator
in steady state in (1+1)D case.

Most prior theoretical work on this topic was done in (1+1)
dimensional spacetimes. However since most experimental
proposals on the detection of Unruh effect are designed for the
physical four dimensional spacetime, it is necessary to do a
detailed analysis for (3+1) dimensions. Although tempting, one
cannot assume that all (3+1) results are equivalent to those from
(1+1) calculations. First, there are new divergences in the (3+1)
case to deal with. Second, the structure of the retarded field in
(3+1) dimensional spacetime is much richer: it consists of a bound
field (generalized Coulomb field) and a radiation field with a
variety of multipole structure, while the (1+1) case has only
the radiation field in a different form. Third, an earlier work
of one of us \cite{lin03b} showed that there is some constant negative
monopole radiation emitted from a detector initially in the ground
state and uniformly accelerated in (3+1)D Minkowski space, and
claimed that this signal could be an evidence of the Unruh effect.
This contradicts the results reported by HR \cite{CapHR} and others
from the (1+1)D calculations. We need to clarify this discrepancy
and determine the cause of it, by studying the complete process from
transient to steady state. In particular, since radiation only
exists under nonequilibrium conditions in the (1+1) case, it is
crucial to understand the transient effects in the (3+1) case to
gauge our expectation of what could be, against what would be,
observed in laboratories.

In conceptual terms, one is tempted to invoke stationarity and
thermality conditions for the description of an UAD. This is indeed
a simple and powerful way to understand its physics if the 
detector undergoes uniform acceleration and interacts with the 
field all throughout (e.g., because of the
stationarity of the problem in the Rindler proper time it is
guaranteed that the total boost energy-operator is conserved).
However, this argument is inapplicable for transient epochs during
which the physics is quite different (see, e.g., the inertial to
uniform acceleration motion treated in \cite{RHK}). Likewise, one
can invoke the thermality condition (i.e., the thermal radiance
experienced by a UAD is equivalent to that of an inertial detector
in a thermal bath) to obtain results based on simple reasonings. But
then we note that the thermality condition does not uniquely arise
from uniform acceleration conditions. For example if the motion is
rapidly altered \cite{HRthermal} 
the radiation produced can be approximately thermal. This thermality
in emitted radiation (e.g., from sudden injection of atoms into a
cavity) is similar to those encountered in cosmological particle
creation \cite{Parker76}, but has a different physical origin from
Unruh effect which is similar to particle creation from black holes
(Hawking effect) \cite{Haw75} (see, e.g., \cite{Scullyetal,
HuRouraPRL}).


In terms of methodology, instead of using the more sophisticated
influence functional method as in the earlier series of papers on
accelerated detectors \cite{RHA,RHK} and moving charges
\cite{JH1,IARD,GH1}, our work here follows more closely the work of
HR who used the Heisenberg operator method to calculate the
two-point function and the stress-energy tensor of a massless quantum
scalar field. In our analysis based on the (3+1)D Unruh-DeWitt detector
theory we found the full and exact dynamics of the detector and the
field in terms of their Heisenberg operator evolution, thus making
available the complete quantum and statistical information for this
detector-field system, enabling us to address the interplay of
thermal radiance in the detector, vacuum polarization cloud around
the detector, quantum fluctuations and radiation, and emitted flux
of classical radiation.

The paper is organized as follows. In Sec.II we introduce the
Unruh-DeWitt detector theory. Then in Sec.III we describe the
quantum dynamics of the detector-field system in the Heisenberg
picture, yielding the expectation values of the detector two-point
function with respect to the Minkowski vacuum and a detector
coherent state in Sec.IV. With these results we derive the two-point
function of the quantum field and describe what constitutes the
``vacuum polarization" around the detector in Sec.V. Then in Sec.VI
we
calculate the quantum expectation values of the stress-energy tensor
induced by the uniformly accelerated detector. This allows us to
explore the conservation law and derive the quantum radiation formula.
A comparison with the results in Ref.\cite{lin03b} follows in Sec.VII.
Finally, we summarize our findings in Sec.VIII.

\section{Unruh-DeWitt Detector Theory}

The total action of the detector-field system is given by
\begin{equation}
  S=S_{Q} +S_\Phi + S_I, \label{Stot1}
\end{equation}
where $Q$ is the internal degree of freedom of the detector,
assumed to be a harmonic oscillator with mass $m_0$ and a (bare)
natural frequency $\Omega_0$:
\begin{equation}
  S_{Q}=\int d\tau {m_0\over 2}\left[ \left(\partial_\tau Q\right)^2
    -\Omega_0^2 Q^2\right].
\end{equation}
Here $\tau$ is the detector's proper time. Henceforth we will use
an overdot on $Q$ to denote $dQ(\tau)/d\tau$. The scalar field $\Phi$
is assumed to be massless,
\begin{equation}
  S_\Phi = -\int d^4 x {1\over 2}\partial_\mu\Phi\partial^\mu\Phi.
\end{equation}
The interaction action $S_I$ for the Unruh-DeWitt (UD) detector theory
has the form \cite{Unr76,DeW79},
\begin{equation} 
  S_I = {\lambda_0} \int d\tau \int d^4 x Q(\tau) \Phi (x)
  \delta^4\left(x^{\mu}-z^{\mu}(\tau)\right). \label{Sint}
\end{equation}
where  $\lambda_0$ is the coupling constant.
This can be regarded as a simplified version of an atom.

Below we will consider the UD detector moving in a prescribed
trajectory $z^\mu(\tau)$ in a four-dimensional Minkowski spacetime
with metric $\eta_{\mu\nu} =$ diag$(-1, 1,1,1)$ and line element
$ds^2 = \eta_{\mu\nu} dx^{\mu}dx^{\nu}$.
By ``prescribed" we mean the trajectory of the detector is not
considered as a dynamical variable, thus we ignore the
backreaction effect of the field on the trajectory.  (See
Ref.\cite{JH1} for an example where the trajectory and the field
are determined self-consistently by each other.)  The detector is
made (by the act of an external agent) to go along the worldline
\begin{equation}
  z^\mu (\tau) = ( a^{-1}\sinh a\tau, a^{-1}\cosh a\tau,0,0),
  \label{cltraj}
\end{equation}
parametrized by its proper time $\tau$. This is the trajectory of a
uniformly accelerated detector situated in Rindler wedge R 
(the portion $t-x^1<0$ and $t+x^1>0$ of Minkowski space; 
see Chapter 4 of Ref.\cite{BD}).

\section{Quantum Theory in Heisenberg Picture}

The conjugate momenta  ($P(\tau), \Pi(x)$) of dynamical variables
($Q(\tau), \Phi(x)$) are defined 
by
\begin{eqnarray}
  P(\tau) &=& {\delta S\over \delta \dot{Q}(\tau)}= m_0 \dot{Q}(\tau),\\
  \Pi(x) &=& {\delta S\over \delta \partial_t\Phi(x)} = \partial_t \Phi(x).
\end{eqnarray}
By treating the above dynamical variables as operators
and introducing the equal time commutation relations,
\begin{eqnarray}
  [ \hat{Q}(\tau), \hat{P}(\tau) ] &=& i\hbar ,\label{QPCM} \\
  \left. [ \hat{\Phi}(t,{\bf x}),\hat{\Pi} (t,{\bf x'}) ]\right.
  &=& i\hbar\delta^3 ({\bf x}-{\bf x'}),\label{phipiCM}
\end{eqnarray}
one can write down the Heisenberg equations of motion for the operators
and obtains
\begin{eqnarray}
  \partial_\tau^2\hat{Q}(\tau)+\Omega_0^2 \hat{Q}(\tau) &=&
  {\lambda_0\over m_0}\hat{\Phi}(\tau,{\bf z}(\tau)),\label{eomq}
 \\
  \left( \partial_t^2-\nabla^2\right)\hat{\Phi}(x) &=& {\lambda_0}
  \int_0^\infty d\tau \hat{Q}(\tau)\delta^4(x-z(\tau)),\label{eomPhi}
\end{eqnarray}
which have the same form as the
classical Euler-Lagrange equations.

Suppose the system is prepared before $\tau=0$, and the coupling $S_I$
is turned on precisely at the moment $\tau=0$ when we allow all
the dynamical variables to begin to interact and evolve under the
influence of each other. (The consequences of this sudden
switch-on and the assumption of a factorizable initial state for
the combined system a quantum Brownian oscillator plus oscillator
bath is described in some details in \cite{HPZ}). By virtue of the
linear coupling $(\ref{Sint})$, the time evolution of
$\hat{\Phi}({\bf x})$ is simply a linear transformation in the
phase space spanned by the orthonormal basis $( \hat{\Phi}({\bf x}),
\hat{\Pi}({\bf x}) , \hat{Q},\hat{P})$, that is,
$\hat{\Phi}(x)$ can be expressed in the form
\begin{equation}
  \hat{\Phi}(t,{\bf x}) = \int d^3 x'\left[ f^\Phi(t,{\bf x},{\bf x'})
  \hat{\Phi}(0,{\bf x'})+f^\Pi(t,{\bf x},{\bf x'})\hat{\Pi}(0,{\bf x'})
  \right]+f^Q(x)\hat{Q}(0)+ f^P(x)\hat{P}(0). \label{defPhi}
\end{equation}
Here $f^\Phi(x,{\bf x'}), f^\Pi(x,{\bf x'}), f^Q(x)$ and
$f^P(x)$ are c-number functions of spacetime. Similarly,
the operator $\hat{Q}(\tau)$ can be written as
\begin{equation}
  \hat{Q}(\tau) = \int d^3 x' \left[q^\Phi(\tau,{\bf x'})
    \hat{\Phi}(0,{\bf x'})+q^\Pi(\tau, {\bf x'})\hat{\Pi}(0,{\bf x'})
    \right]+ q^Q(\tau) \hat{Q}(0)+ q^P(\tau) \hat{P}(0), \label{defq}
\end{equation}
with c-number functions $q^Q(\tau), q^P(\tau),
q^\Phi(\tau,{\bf x'})$ and $q^\Pi(\tau,{\bf x'})$.

For the case with initial operaters being the free field operators,
namely, $\hat{\Phi} (0,{\bf x})=\hat{\Phi}_0({\bf x})$,
$\hat{\Phi}(0,{\bf x})=\hat{\Pi}_0({\bf x})$,
$\hat{Q}(0)=\hat{Q}_0$ and $\hat{P}(0)=\hat{P}_0$,
one can go further by introducing the complex operators
$\hat{b}_{\bf k}$ and $\hat{a}$:
\begin{eqnarray}
  \hat{\Phi}_0({\bf x}) &=& \int {d^3k\over (2\pi)^3}
    \sqrt{\hbar\over 2\omega}\left[ e^{i{\bf k\cdot x}}\hat{b}_{\bf k}
    + e^{-i{\bf k\cdot x}}\hat{b}^\dagger_{\rm k}\right] ,\\
  \hat{\Pi}_0({\bf x}) &=& \int {d^3k\over (2\pi)^3}
    \sqrt{\hbar\over 2\omega}(-i\omega )\left[ e^{i{\bf k\cdot x}}
    \hat{b}_{\bf k}-e^{-i{\bf k\cdot x}}\hat{b}^\dagger_{\bf k}\right]
\end{eqnarray}
with $\omega \equiv |{\bf k}|$, and
\begin{equation}
  \hat{Q}_0 = \sqrt{\hbar\over 2\Omega_r m_0}(\hat{a}+\hat{a}^\dagger),
  \;\;\;\;\;
  \hat{P}_0 = -i\sqrt{\hbar\Omega_rm_0\over 2}( \hat{a}-\hat{a}^\dagger ).
\end{equation}
Note that, instead of $\Omega_0$, we use the renormalizd natural frequency
$\Omega_r$ (to be defined in $(\ref{renormO})$) in the definition of
$\hat{a}$. Then the commutation relations $(\ref{QPCM})$ and
$(\ref{phipiCM})$ give
\begin{equation}
  [ \hat{a},\hat{a}^\dagger ]=1, \;\;\;\;\;
  [ \hat{b}_{\bf k}, \hat{b}_{\bf k'}^\dagger ] =
    (2\pi)^3\delta^3({\bf k}-{\bf k'}),
\end{equation}
and the expressions $(\ref{defPhi})$ and $(\ref{defq})$ can be re-written as
\begin{eqnarray}
  \hat{\Phi}(t,{\bf x}) &=&\hat{\Phi}_b(x) +\hat{\Phi}_a(x),\label{phiab}\\
  \hat{Q}(\tau) &=& \hat{Q}_b(\tau) + \hat{Q}_a(\tau)  \label{qab}
\end{eqnarray}
where
\begin{eqnarray}
  \hat{\Phi}_b(x) &=& \int {d^3k\over (2\pi)^3}\sqrt{\hbar\over 2\omega}
    \left[f^{(+)}(t,{\bf x};{\bf k})\hat{b}_{\bf k} +
    f^{(-)}(t,{\bf x};{\bf k})\hat{b}_{\bf k}^\dagger
    \right],\label{Phiv}\\
  \hat{\Phi}_a(x) &=& \sqrt{\hbar \over 2\Omega_r m_0}\left[f^a(t,{\bf x})
    \hat{a}+ f^{a*}(t,{\bf x})\hat{a}^\dagger \right], \label{PhiB}\\
  \hat{Q}_b(\tau) &=& \int {d^3 k\over (2\pi)^3}\sqrt{\hbar\over 2\omega}
  \left[q^{(+)}(\tau,{\bf k})\hat{b}_{\bf k} +
    q^{(-)}(\tau,{\bf k})\hat{b}_{\bf k}^\dagger\right],\label{Qv} \\
  \hat{Q}_a(\tau) &=& \sqrt{\hbar\over 2\Omega_r m_0}\left[
    q^a(\tau)\hat{a}+q^{a*}(\tau)\hat{a}^\dagger \right] .\label{QB}
\end{eqnarray}
The whole problem therefore can be transformed to solving
c-number functions $f^s(x)$ and $q^s(\tau)$ from $(\ref{eomq})$
and $(\ref{eomPhi})$ with suitable initial conditions.
Since $\hat{Q}$ and $\hat{\Phi}$ are hermitian, one has $f^{(-)}=
(f^{(+)})^*$ and $q^{(-)}= (q^{(+)})^*$. Hence it is sufficient
to solve the c-number functions $f^{(+)}(t,{\bf x};{\bf k})$,
$q^{(+)}(\tau,{\bf k})$, $f^a(t,{\bf x})$ and $q^a(\tau)$.
To place this in a more general setting, let us perform a Lorentz
transformation shifting $\tau =0$ to $\tau=\tau_0$, and define
\begin{equation}
  \eta \equiv \tau -\tau_0.
\end{equation}
This does not add any complication to our calculation.
Now the coupling between the detector and the field would be turned on
at $\tau=\tau_0$. We are looking for solutions with the initial conditions
\begin{eqnarray}
 && f^{(+)}(t(\tau_0),{\bf x};{\bf k}) = e^{i{\bf k\cdot x}},\;\;\;
    \partial_t f^{(+)}(t(\tau_0),{\bf x};{\bf k})=
      -i\omega e^{i{\bf k\cdot x}}, \;\;\;
q^{(+)}(\tau_0;{\bf k})= \dot{q}^{(+)}(\tau_0;{\bf k}) = 0, \label{IC+}\\
 && f^a (t(\tau_0),{\bf x}) =\partial_t f^a (t(\tau_0),{\bf x}) =0,
 \;\;\; q^a(\tau_0) = 1, \;\;\; \dot{q}^a(\tau_0)= -i\Omega_r. \label{ICa}
\end{eqnarray}

\subsection{Solving for $f^{(+)}$ and $q^{(+)}$}
\label{solvefq}

The method to solve for $f$ and $q$ are analogous to what we did
in classical field theory. First, we find an expression relating
the harmonic oscillator and the field amplitude right
at the detector. Then substituting this relation into the
equation of motion for the oscillator, we obtain a complete
equation of motion for $q$ with full information of the field.
Last, we solve this complete equation of motion for $q$, and from
its solution determine the field $f$ consistently.

Eq.$(\ref{eomPhi})$ implies that
\begin{equation}
  (\partial_t^2-\nabla^2)f^{(+)}(x;{\bf k}) =
    {\lambda_0}\int_{\tau_0}^\infty d\tau
    \delta^4(x-z(\tau))q^{(+)}(\tau;{\bf k}).
\label{eomPhi+}
\end{equation}
The general solution for $f^{(+)}$ reads
\begin{equation}
  f^{(+)}(x;{\bf k}) = f_0^{(+)}(x;{\bf k})
  + f_1^{(+)}(x;{\bf k}), \label{Phi+}
\end{equation}
where
\begin{equation}
  f_0^{(+)}(x;{\bf k})\equiv  e^{-i\omega t+i{\bf k\cdot x}}
  \label{freephi}
\end{equation}
is the free field solution, and
\begin{equation}
  f_1^{(+)}(x;{\bf k})\equiv {\lambda_0} \int_{\tau_0}^\infty d\tau
    G_{\rm ret}(x;z(\tau))q^{(+)}(\tau;{\bf k})
\label{fretard}
\end{equation}
is the retarded solution, which looks like the retarded field in classical
field theory. Here $\omega=|{\bf k}|$ and the retarded Green's function
$G_{\rm ret}$ in Minkowski space is given by
\begin{equation}
  G_{\rm ret}(x,x') = {1\over 4\pi}\delta(\sigma)\theta(t-t') \label{Gret}
\end{equation}
with $\sigma\equiv -(x_\mu-x'_\mu)(x^\mu-x'^\mu)/2$. Applying the explicit
form of the retarded Green's function, one can go further to write
\begin{equation}
  f_1^{(+)}(x;{\bf k}) =
    {\lambda_0\theta(\eta_-)  \over 2\pi a X}q^{(+)}(\tau_-;{\bf k}),
    \label{Phi+1}
\end{equation}
where
\begin{eqnarray}
  X &\equiv& \sqrt{(-UV+\rho^2+a^{-2})^2+4a^{-2}UV}, \label{defX}\\
  \tau_- &\equiv& -{1\over a}\ln {a\over 2|V|}\left(X-UV+\rho^2+a^{-2}
  \right),\label{deftaum}\\
  \eta_- &\equiv& \tau_- -\tau_0,
\end{eqnarray}
with $\rho \equiv \sqrt{x_2{}^2 +x_3{}^2}$, $U\equiv t-x^1$ and
$V \equiv t+x^1$.

The formal retarded solution $(\ref{Phi+1})$ is singular on the
trajectory of the detector. To deal with the singularity, note that 
the UD detector here is a quantum mechanical object, and the detector 
number would always be one. This means that at the energy threshold 
of detector creations, there is a natural cutoff on frequency, 
which sets an upper bound on the resolution to be explored in our theory.
Thus it is justified to assume here that the detector has a finite extent
$O(\Lambda^{-1})$, which will introduce the back reaction on the detector.

Let us regularize the retarded Green's function by invoking the
essence of effective field theory:
\begin{equation}
  G^\Lambda_{\rm ret}(x,x') = 
  {1\over 4\pi}\sqrt{8\over \pi}\Lambda^2 e^{-2\Lambda^4\sigma^2}
  \theta(t-t') . \label{GretL}
\end{equation}
(For more details on this regularization scheme, see Refs.\cite{JH1,GH1}.)
 With this, right on the trajectory,
the retarded solution for large $\Lambda$ is
\begin{equation}
  f^{(+)}_1(z(\tau);{\bf k}) =
  {\lambda_0\over 4\pi}\left[
  \Lambda \zeta q^{(+)}(\tau;{\bf k})
  -\partial_\tau q^{(+)}(\tau;{\bf k}) + O(\Lambda^{-1})\right],
\end{equation}
where $\zeta = 2^{7/4} \Gamma ( 5/4 )/\sqrt{\pi}$.
Substituting the above expansion into $(\ref{eomq})$ and neglecting
$O(\Lambda^{-1})$ terms, one obtains the equation of motion for
$q^{(+)}$ with back reaction,
\begin{equation}
  (\partial_\tau^2 +2\gamma\partial_\tau+ \Omega_r^2)q^{(+)}(\tau;{\bf k}) =
  {\lambda_0\over m_0} f^{(+)}_0(z(\tau);{\bf k}).\label{eomq2}
\end{equation}
Fortunately, there is no higher derivatives of $q$ present in the
above equation of motion. Now $q^{(+)}$ behaves like a damped harmonic
oscillator driven by the vacuum fluctuations of the scalar field, with
the damping constant
\begin{equation}
  \gamma \equiv {\lambda_0^2\over 8\pi m_0},
\end{equation}
and the renormalized natural frequency
\begin{equation}
  \Omega_r^2 \equiv \Omega_0^2-{\lambda_0^2\Lambda\zeta \over 4\pi m_0}.
  \label{renormO}
\end{equation}

In $(\ref{eomq2})$, the solution for $q^{(+)}$ compatible with the initial
conditions $q^{(+)}(\tau_0;{\bf k})=\dot{q}^{(+)}(\tau_0;{\bf k})=0$ is
\begin{equation}
  q^{(+)}(\tau;{\bf k}) = {\lambda_0\over m_0 }\sum_{j=+,-}
  \int^\tau_{\tau_0} d\tau' c_j e^{w_j(\tau-\tau')}
  f^{(+)}_0(z(\tau');{\bf k}),\label{q+1}
\end{equation}
where $f^{(+)}_0$ has been given in $(\ref{freephi})$, $c_\pm$ and $w_\pm$
are defined as
\begin{equation}
  c_\pm = \pm {1\over 2i\Omega}, \;\;\;
  w_\pm = -\gamma \pm i\Omega,
\end{equation}
with
\begin{equation}
  \Omega \equiv \sqrt{\Omega_r^2 -\gamma^2}. \label{bigW}
\end{equation}
Throughout this paper we consider only the under-damped case with
$\gamma^2 < \Omega_r^2$, so $\Omega$ is always real.

\subsection{Solving for $f^a$ and $q^a$}

Similarly, from $(\ref{eomq})$, $(\ref{eomPhi})$, $(\ref{phiab})$
and $(\ref{qab})$, the equations of motion for $f^a$ and $q^a$ read
\begin{eqnarray}
  &&   \left( \partial_t^2-\nabla^2\right)f^a(x)= \lambda_0\int d\tau
    \delta^4(x-z(\tau))q^a(\tau),\\
  && (\partial_\tau^2 +\Omega_0^2)q^a (\tau) = {\lambda_0\over m_0}
    f^a (z(\tau)). \label{eomqb}
\end{eqnarray}
The general solutions for $f^a$, similar to $(\ref{Phi+})$, is
\begin{equation}
  f^a(x) = f_0^a(x) +{\lambda_0}\int_{\tau_0}^\infty d\tau
  G_{\rm ret}(x;z(\tau)) q^a(\tau_-) \label{Phib}
\end{equation}
However, according to the initial condition $(\ref{ICa})$,
one has $f^a_0 = 0$, hence
\begin{equation}
  f^a(x)= {\lambda_0\theta(\eta_-)  \over 2\pi a X}q^a(\tau_-).
  \label{Phibfinal}
\end{equation}
Again, the value of $f^a$ is singular right at the position of
the detector. Performing the same regularization as those for
$q^{(+)}$, $(\ref{eomqb})$ becomes (cf. $(\ref{eomq2})$)
\begin{equation}
 \left(\partial_\tau^2 + 2\gamma \partial_\tau + \Omega_r^2\right)
 q^a (\tau) = 0, \label{noforceeom}
\end{equation}
which describes a damped harmonic oscillator free of driving force.
The solution consistent with the initial condition $q^a(\tau_0)=1$ and
$\dot{q}^b(\tau_0)= -i\Omega_r$ reads
\begin{equation}
  q^a(\tau) = {1\over 2}\theta(\eta)e^{-\gamma\eta}
  \left[\left(1-{\Omega_r+ i\gamma\over\Omega}\right)e^{i\Omega\eta} +
  \left(1+{\Omega_r+ i\gamma\over\Omega}\right)e^{-i\Omega\eta}\right].
\label{qb}
\end{equation}

\section{Two-Point Functions of the Detector}

As shown in the previous section, as $\hat{Q}$ evolves,  some
non-zero terms proportional to $\hat{\Phi}$ and $\hat{\Pi}$ will
be generated. Suppose the detector is initially prepared in a
state that can be factorized into the quantum state
$\left|\right.q\left. \right>$ for $Q$ and the Minkowski vacuum
$|\left. 0_M\right> $ for the scalar field $\Phi$, that is,
\begin{equation}
  | \tau_0 \left.\right> =  \left|\right.q\left. \right> |\left. 0_M\right>
\label{initQS}
\end{equation}
then the two-point function of $Q$ will split into two parts,
\begin{eqnarray}
  \left<\right. Q(\tau)Q(\tau')\left.\right>  &=&  \left<\right. 0_M|
  \left< \right. q\left.\right| \left[\hat{Q}_b(\tau) + \hat{Q}_a(\tau)\right]
  \left[\hat{Q}_b(\tau) + \hat{Q}_a(\tau)\right]\left|\right.q\left. \right>
  |0_M\left.\right> \nonumber\\ &=& \left<\right.q \,|\, q\left.\right>
  \left< \right. Q(\tau)Q(\tau')\left.\right>_{\rm v}+ \left< \right.
  Q(\tau)Q(\tau')\left.\right>_{\rm a}\left< 0_M| 0_M\right>. \label{splitQQ}
\end{eqnarray}
where, from $(\ref{qab})$,
\begin{eqnarray}
  \left< \right. Q(\tau)Q(\tau')\left.\right>_{\rm v} &=&  \left< 0_M\right.|
    \hat{Q}_b(\tau)\hat{Q}_b(\tau')|\left. 0_M\right>, \label{defQQv}\\
  \left< \right. Q(\tau)Q(\tau')\left.\right>_{\rm a} &=&
  \left< \right. q\left.\right|\hat{Q}_a(\tau)\hat{Q}_a(\tau)
  \left|\right.q\left. \right>. \label{defQQb}
\end{eqnarray}
Similar splitting happens for every two-point function of
$\hat{\Phi}(x)$ as well as for the stress-energy tensor.

Observe that $\left< \right. Q(\tau)Q(\tau')\left.\right>_{\rm
v}$ depends on the initial state of the field, or the Minkowski
vacuum, while $\left< \right. Q(\tau)Q(\tau') \left.\right>_{\rm
a}$ depends on the initial state of the detector only. One can
thus interpret  $\left< \right. Q(\tau)Q(\tau')\left.
\right>_{\rm v} $ as accounting for the response to the vacuum
fluctuations, while $\left< \right. Q(\tau)Q(\tau')\left.
\right>_{\rm a}$ corresponds to the intrinsic quantum
fluctuations in the detector.

In the following, we will demonstrate the explicit forms of some two-point
functions we have obtained and analyze their behavior. To distinguish the
quantum or classical natures of these quantities, the initial quantum state
$\left|\right. q\left.\right>$ will be taken to be the coherent state
\cite{scully},
\begin{equation}
  \left|\right. q\left.\right> = e^{-\alpha^2/2}\sum_{n=0}^\infty
  {\alpha^n\over\sqrt{n!}}\left|\right. n\left.\right>,\label{cohere}
\end{equation}
where $\left|\right. n\left.\right>$ is the $n$-th excited state for the
free detector, and $\alpha = q_0\sqrt{\Omega_r/ 2\hbar}$ with a constant
$q_0$. The representation of $\left|\right. q\left.\right>$ in $Q$-space
reads
\begin{equation}
  \psi(Q,\tau_0) = \left({\Omega_r\over\pi\hbar}\right)^{1/4}
  e^{-\Omega_r(Q-q_0)^2/2\hbar},
\end{equation}
which is a wave-packet centered at $q_0$ with the spread identical to the
one for the ground state.

\subsection{Expectation value of the detector two-point function with
respect to the Minkowski vacuum}

Along the trajectory $z^\mu(\tau)$ in $(\ref{cltraj})$, performing a
Fourier transformation with respect to $\tau$ on $(\ref{freephi})$,
one has
\begin{equation}
  f_0^{(+)}(z(\tau);{\bf k}) \equiv \int d\kappa e^{-i\kappa\tau}
  \varphi(\kappa,{\bf k}), \label{f0+zt}
\end{equation}
where the frequency spectrum of the Minkowski mode from the viewpoint
of the UAD,
\begin{equation}
  \varphi(\kappa, {\bf k}) = {e^{-\pi \kappa/2a}\over \pi a}
  \left( \omega-k_1\over \omega+k_1\right)^{-{i\kappa\over 2a}}
  K_{-{i\kappa\over a}}\left(\sqrt{k_2{}^2+k_3{}^2}/a\right),
\label{rindspec}
\end{equation}
is not trivial any more. Given the result of the integration,
\begin{eqnarray}
  \int {\hbar d^3 k\over (2\pi)^3 2\omega}  \varphi(\kappa,{\bf k})
  \varphi^*(\kappa',{\bf k}) &=&{\hbar\over (2\pi)^2}
  {\kappa\over 1-e^{-2\pi \kappa/a}}\delta(\kappa-\kappa'),\label{trick1}
\end{eqnarray}
a Planck factor with the Unruh temperature $a/2\pi$ emerges. Then 
from $(\ref{defQQv})$, $(\ref{q+1})$ and $(\ref{f0+zt})$, one has
\begin{eqnarray}
  \left<\right. Q(\eta)Q(\eta')\left.\right>_{\rm v} &=& \hbar\int 
  {d^3 k\over(2\pi)^3 2\omega} q^{(+)}(\tau;{\bf k}) q^{(-)}(\tau;{\bf k})
  \label{QQvorigin}\\
  &=& {\lambda_0^2 \hbar\over (2\pi)^2 m_0^2 }\sum_{j,j'} \int 
  {\kappa d\kappa\over 1-e^{-2\pi \kappa/a}} {c_j c^*_{j'} 
  e^{-i\kappa(\tau_0-\tau'_0)}\over (w_j+i\kappa)(w^*_{j'}-i \kappa)} 
  \times \nonumber\\ & & 
  \left[ e^{w_j(\tau-\tau_0)}- e^{-i\kappa(\tau-\tau_0)}\right]
  \left[ e^{w^*_{j'}(\tau'-\tau'_0)}-e^{i\kappa(\tau'-\tau'_0)}\right],
\end{eqnarray}
where the integrand has poles at $\kappa = \pm\Omega -i\gamma$ and 
$\kappa = \pm ina$, $n\in N$.
Let $\tau'_0 <\tau_0 < \tau< \tau'$ and taking the coincidence limit, 
one obtains
\begin{eqnarray}
  \left<\right. Q(\eta)^2\left.\right>_{\rm v} &=&  \lim_{\eta'\to\eta}
  {1\over 2}\left<\right. \left\{ Q(\eta),Q(\eta')\right\}
  \left.\right>_{\rm v}\nonumber\\
  &=& {\hbar\lambda_0^2\over (2\pi m_0 \Omega)^2} \theta(\eta)
   {\rm Re}\left\{
  (\Lambda_0-\ln a) e^{-2\gamma\eta}\sin^2\Omega\eta \right. \nonumber\\ 
  &+& {a\over 2}e^{-(\gamma +a)\eta}\left[{F_{\gamma+i\Omega}(e^{-a\eta})
\over \gamma+i\Omega+a}\left( -{i\Omega\over\gamma}\right)e^{-i\Omega\eta}
   + {F_{-\gamma - i\Omega}(e^{-a\eta})\over \gamma + i\Omega-a}\left(
    \left(1+{i\Omega\over\gamma}\right)e^{i\Omega\eta} 
    -e^{-i\Omega\eta}\right)\right]\nonumber\\
  &-& {1\over 4}\left[ \left({i\Omega\over\gamma}+ e^{-2\gamma\eta}
  \left({i\Omega\over\gamma}+1 -e^{-2i\Omega\eta}\right)\right)
  \left( \psi_{\gamma+i\Omega}+ \psi_{-\gamma-i\Omega}\right)\right.
  \nonumber\\ & & \left.\left. - \left(-{i\Omega\over\gamma}+
  e^{-2\gamma\eta}\left({i\Omega\over\gamma}+1 -e^{-2i\Omega\eta}
  \right)\right)i\pi\coth{\pi\over a}(\Omega-i\gamma)\right]\right\}.
\label{Q^2v}
\end{eqnarray}
Here $F_s(t)$ is defined by the hyper-geometric function as
\begin{equation}
  F_s(t) \equiv {}_2 F_1\left(1+{s\over a},1,2+{s\over a}; t\right),
\end{equation}
and
\begin{equation}
  \psi_s \equiv \psi\left(1+{s\over a}\right)
\end{equation}
is the poly-gamma function. The divergent $\Lambda_0$-term is produced 
by the coincidence limit: as $\eta' \to \eta$, $\Lambda_0 \to -
\gamma_{e}-\ln |\tau'_0-\tau_0|$ with the Euler's constant $\gamma_e$. 
Since $|\tau'_0-\tau_0|$ characterizes the time scale that the 
interaction is turned on, $\Lambda_0$ could be finite in real processes. 
In any case, for every finite value of $\Lambda_0$, 
the first line of the result in $(\ref{Q^2v})$ vanishes as 
$\eta\to\infty$.

\begin{figure}
\includegraphics[width=8cm]{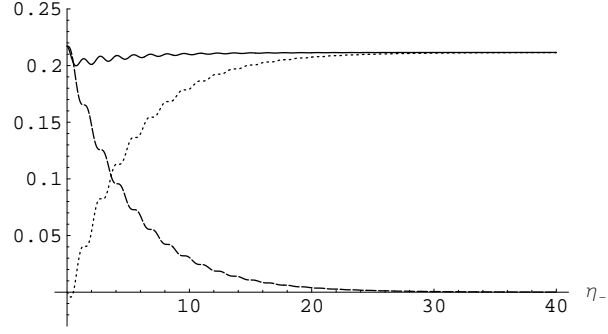}
\caption{Plot of $\left<\right. Q(\eta)^2\left.\right>_{\rm v}$
(dotted line, Eq.($\ref{Q^2v}$) with $\Lambda_0$-term excluded),
$\left<\right. Q(\eta)^2\left.\right>_{\rm a}^{\rm qm}$ (dashed line,
Eq.($\ref{q2bqm}$)), and the sum of these two ($\left<\right. 
\Delta Q(\eta)^2\left.\right>$, solid line). Here
we have taken $a=1$, $\gamma=0.1$, $\Omega=2.3$ and $m_0=1$. }
\label{QQvseta}
\end{figure}

In FIG. \ref{QQvseta}, we show the $\left<\right.Q(\eta)^2\left.
\right>_{\rm v}$ without $\Lambda_0$-term in dotted line.
Roughly speaking the curve saturates exponentially in the
detector's proper time. As $\eta\to\infty$, $\left<\right.
Q(\eta)^2\left.\right>_{\rm v}$ saturates to the value
\begin{equation}
  \lim_{\eta\to\infty}\left<\right. Q(\eta)^2\left.\right>_{\rm v} =
  {\hbar\over 2\pi m_0\Omega} {\rm Re}\, \left[
  {ia\over \gamma+i\Omega}-2i\psi_{\gamma+i\Omega}\right].\label{QQsat}
\end{equation}
For $\gamma< a$, the time scale of the rise is about $1/2\gamma$,
which can be read off from the $e^{-2\gamma\eta}$ in
$(\ref{Q^2v})$. From there one can also see that the small
oscillation around the rising curve has a frequency of
$O(\Omega)$.

For $\left<\right. Q(\eta)\dot{Q}(\eta)\left.\right>_{\rm v}$,
it will be clear that what is interesting for the calculation of
the flux is the combined quantities like $\left<\right.
Q(\eta)\dot{Q}(\eta)\left.\right>_{\rm v} +\left<\right.
\dot{Q}(\eta) Q(\eta)\left.\right>_{\rm v}$. Notice that
\begin{equation}
  \left<\right. Q(\eta)\dot{Q}(\eta)\left.\right>_{\rm v} +
  \left<\right. \dot{Q}(\eta)Q(\eta)\left.\right>_{\rm v} =
  \partial_\tau \left<\right. Q(\eta)^2\left.\right>_{\rm v}.
\end{equation}
With the result of $\left<\right. Q(\eta)^2\left.\right>_{\rm v}$, this
calculation is straightforward.
Let us turn to the two-point functions of $\dot{Q}$. Similar to
$\left<\right. Q(\eta)^2\left.\right>_{\rm v}$, one has
\begin{eqnarray}
\left<\right. \dot{Q}(\eta)\dot{Q}(\eta')\left.\right>_{\rm v} &=&
  \int {\hbar d^3 k\over (2\pi)^3 2\omega} \dot{q}^{(+)}(\tau;{\bf k})
    \dot{q}^{(-)}(\tau;{\bf k})\\
  &=& {\lambda_0^2\hbar\over (2\pi)^2 m_0^2 }\sum_{j,j'} \int
  {\kappa d\kappa \over 1-e^{-2\pi \kappa/a}}{c_j c^*_{j'}
  e^{-i\kappa(\tau_0-\tau'_0)}\over (w_j+i\kappa)(w^*_{j'}-i \kappa)}
  \times \nonumber\\ & &
  \left[ w_j e^{w_j(\tau-\tau_0)}+i\kappa e^{-i\kappa(\tau-\tau_0)}\right]
  \left[ w^*_{j'}e^{w^*_{j'}(\tau'-\tau'_0)}-
  i\kappa e^{i\kappa(\tau'-\tau'_0)}\right]     \label{QtQt'}
\end{eqnarray}
from $(\ref{q+1})$, $(\ref{f0+zt})$ and $(\ref{trick1})$. The coincidence limit
of the above two-point function reads
\begin{eqnarray}
  \left<\right. \dot{Q}(\eta)^2\left.\right>_{\rm v}
    &=& {\hbar\lambda_0^2\over ( 2\pi m_0\Omega)^2}
  \theta(\eta){\rm Re}\,\left\{
   (\Lambda_1-\ln a)\Omega^2 + (\Lambda_0-\ln a)e^{-2\gamma\eta}\left(\Omega
   \cos\Omega\eta-\gamma \sin\Omega\eta \right)^2 
   \right. \nonumber\\
  &+& {a\over 2}(\gamma+i\Omega)^2 e^{-(\gamma +a)\eta} \left[{F_{\gamma+i\Omega}
    (e^{-a\eta}) \over \gamma+i\Omega+a}\left( {i\Omega\over\gamma}\right)
    e^{-i\Omega\eta} + {F_{-\gamma-i\Omega}(e^{-a\eta})\over \gamma +i\Omega-a}\left(
    \left(1-{i\Omega\over\gamma}\right)e^{i\Omega\eta} -e^{-i\Omega\eta}\right)
    \right]\nonumber\\
  &+& {1\over 4} (\gamma +i\Omega)^2\left[\left( {i\Omega\over\gamma}+
    e^{-2\gamma\eta}\left({i\Omega\over\gamma}-1+e^{-2i\Omega\eta}\right)\right)
  \left(\psi_{\gamma+i\Omega}+  \psi_{-\gamma-i\Omega}\right)\right.\nonumber\\
  & & \left.\left. -\left(-{i\Omega\over\gamma}+e^{-2\gamma\eta}
  \left({i\Omega\over\gamma}-1+e^{-2i\Omega\eta}\right)\right)i\pi\coth
  {\pi\over a}(\Omega-i\gamma)\right]\right\}, \label{dotQ^2v}
\end{eqnarray}
where $\Lambda_1 = -\gamma_e -\lim_{\tau'\to\tau}\ln |\tau-\tau'|$ can be
subtracted safely. This will be justified later.

\begin{figure}
\includegraphics[width=8cm]{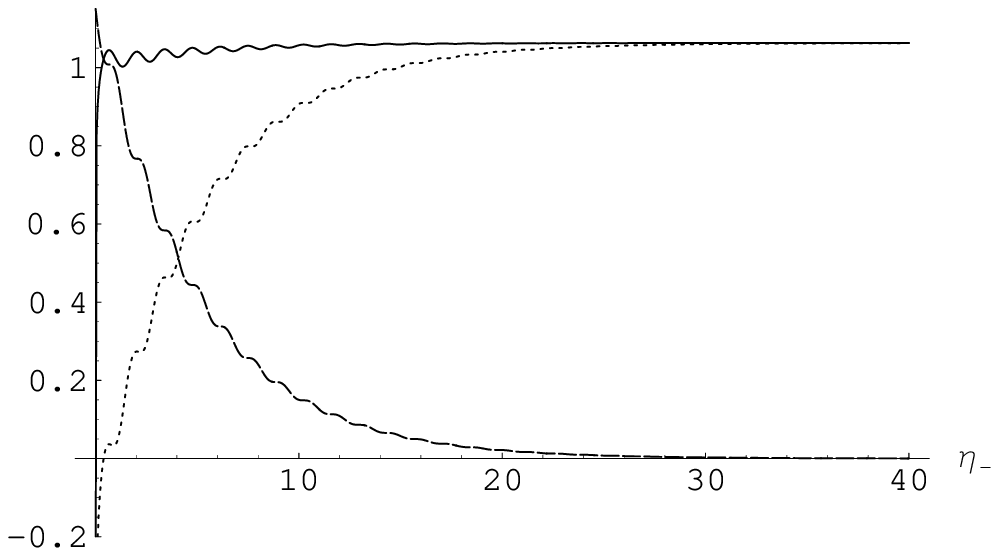}
\caption{Plots of $\left<\right.\dot{Q}(\eta)^2\left.\right>_{\rm v}$
(dotted line, Eq.($\ref{dotQ^2v}$)), $\left<\right.\dot{Q}(\eta)^2\left.
\right>_{\rm a}^{\rm qm}$ (dashed line, Eq.($\ref{dotQ2aqm}$)) and their
sum ($\left<\right. \Delta \dot{Q}(\eta)^2\left.\right>$, solid line).}
\label{PPvseta}
\end{figure}

The subtracted $\left<\right. \dot{Q}(\eta)^2\left.\right>_{\rm
v}$ is illustrated in FIG. \ref{PPvseta} (in which $\Lambda_0$-term has
also been excluded). One can immediately recognize that
$\left<\right.\dot{Q}(\eta)^2 \left.\right>_{\rm v} \sim \ln\eta$
when $\eta$ approaches zero; a new divergence
occurs at $\eta =0$. Mathematically, this logarithmic divergence
comes about because the divergences in the hyper-geometric
functions in $(\ref{dotQ^2v})$ do not cancel each other, unlike
in $\left<\right.Q(\eta)^2\left.\right>_{\rm v}$.
Physically, this divergence at the initial time $\tau_0$ could be
another consequence of the sudden switch-on at $\tau=\tau_0$ or $\eta=0$.
We expect that these ill-behaviors at the start
could be tamed if we turn on the coupling adiabatically. (See
\cite{HPZ} for a discussion on this issue.)

For large $\eta$, the behavior of the dotted curve in FIG. \ref{PPvseta}
is quite similar to the one in FIG. \ref{QQvseta} for $\left<\right.
Q(\tau)^2\left.\right>_{\rm v}$. It saturates to
\begin{equation}
  \lim_{\eta\to\infty}\left<\right. \dot{Q}(\eta)^2\left.\right>_{\rm v}
  ={\hbar\over 2\pi m_0\Omega} {\rm Re}\,\left\{ (\Omega-i\gamma)^2
  \left[ {ia\over \gamma+i\Omega}-2i\psi_{\gamma+i\Omega}\right]
  - 4\gamma \Omega \ln a \right\}.
 \label{PPsat}
\end{equation}
Comparing $(\ref{dotQ^2v})$ and $(\ref{Q^2v})$, their time scales
of saturation ($1/2\gamma$ for $\gamma<a$) and the frequency of
the small ripples on the rising curve ($O(\Omega)$) are also the same.
Note that when $\gamma \ll 1$ and $a$ is finite, $(\ref{PPsat})$ and
$(\ref{QQsat})$ implies
\begin{equation}
  \left<\right. \dot{Q}(\infty)^2\left.\right>_{\rm v} \approx
    \Omega^2\left<\right. Q(\infty)^2\left.\right>_{\rm v},
\end{equation}
which justifies the subtraction of $\Lambda_1$-term in $(\ref{dotQ^2v})$.

\subsection{Expectation values of the detector two-point functions with
respect to a coherent state}

We now derive the expectation values of the detector two-point functions
with respect to the coherent state $(\ref{cohere})$. Subsituting $(\ref{QB})$
into $(\ref{defQQb})$ and using $(\ref{qb})$ and $(\ref{cohere})$, one finds
that
\begin{equation}
  \left<\right.Q(\tau)Q(\tau')\left.\right>_{\rm a} =
  \left<\right.Q(\tau)Q(\tau')\left.\right>_{\rm a}^{\rm qm} +
  \left<\right.Q(\tau)Q(\tau')\left.\right>_{\rm a}^{\rm cl}, \label{QQBgen}
\end{equation}
where
\begin{eqnarray}
  \left<\right.Q(\tau)Q(\tau')\left.\right>_{\rm a}^{\rm qm} &\equiv&
    {\hbar\over 2\Omega_r m_0} q^a(\tau)q^{a*}(\tau'),\\
  \left<\right.Q(\tau)Q(\tau')\left.\right>_{\rm a}^{\rm cl} &\equiv&
{q_0^2\over m_0}{\rm Re}\left[q^a(\tau)\right]{\rm Re}\left[q^a(\tau')\right]
   = \bar{Q}(\tau)\bar{Q}(\tau'),\label{barQ}
\end{eqnarray}
with the mean value
\begin{equation}
  \bar{Q}(\tau)\equiv \left<\right.Q(\tau)\left.\right> =
  {q_0\over\sqrt{m_0}}\theta(\eta)e^{-\gamma\eta}
  \left(\cos\Omega\eta+{\gamma\over\Omega}\sin\Omega\eta\right).
\end{equation}
While the ``qm" term is of purely quantum nature, the ``cl" term is
of classical nature: $\bar{Q}$ is real and $\left<\right.Q(\tau)
Q(\tau')\left.\right>_{\rm a}^{\rm cl}$ does not involve $\hbar$.
Thus the ``cl" term is identified as the semiclassical part of the 
two-point functions. The coincidence limits of the above two-point 
functions are 
\begin{equation}
  \left<\right.Q(\eta)^2\left.\right>_{\rm a}^{\rm qm} =
  {\hbar \theta(\eta)\over 2\Omega^2\Omega_r m_0}e^{-2\gamma\eta}
  \left[\Omega_r^2 - \gamma^2\cos 2\Omega\eta +
  \gamma\Omega\sin 2\Omega\eta\right],\label{q2bqm}
\end{equation}
and $\left<\right.Q(\eta)^2\left.\right>_{\rm a}^{\rm cl} =\bar{Q}(\eta)^2$.

Similarly, it is easy to find
$\left<\right.\dot{Q}(\tau)\dot{Q}(\tau')\left.\right>_{\rm a} =
\left<\right.\dot{Q}(\tau)\dot{Q}(\tau')\left.\right>_{\rm a}^{\rm qm} + \left<\right.\dot{Q}(\tau)\dot{Q}(\tau')\left.\right>_{\rm a}^{\rm cl}$:
\begin{eqnarray}
  \left<\right.\dot{Q}(\tau)\dot{Q}(\tau')\left.\right>_{\rm a}^{\rm qm}
  &=& {\hbar\over 2\Omega_r m_0}\dot{q}^b(\tau)\dot{q}^{b*}(\tau'), \\
  \left<\right.\dot{Q}(\tau)\dot{Q}(\tau')\left.\right>_{\rm a}^{\rm cl}
  &=&
  \dot{\bar{Q}}(\tau)\dot{\bar{Q}}(\tau'),
\end{eqnarray}
and their coincidence limit,
\begin{equation}
  \left<\right.\dot{Q}(\eta)^2\left.\right>_{\rm a}^{\rm qm} =
  {\hbar \Omega_r \over 2\Omega^2 m_0}\theta(\eta)e^{-2\gamma\eta}\left[
  \Omega_r^2 -\gamma^2\cos 2\Omega\eta -\gamma\Omega\sin 2\Omega\eta\right],
  \label{dotQ2aqm}
\end{equation}
and $\left<\right.\dot{Q}(\eta)^2\left.\right>_{\rm a}^{\rm cl}
=\dot{\bar{Q}}(\eta)^2$. Also one has
$\left<\right. Q(\tau)\dot{Q}(\tau)\left.\right>_{\rm a} +
  \left<\right. \dot{Q}(\tau)Q(\tau)\left.\right>_{\rm a} =
  \partial_\tau \left<\right. Q(\tau)^2\left.\right>_{\rm a}$.

\begin{figure}
\includegraphics[width=8cm]{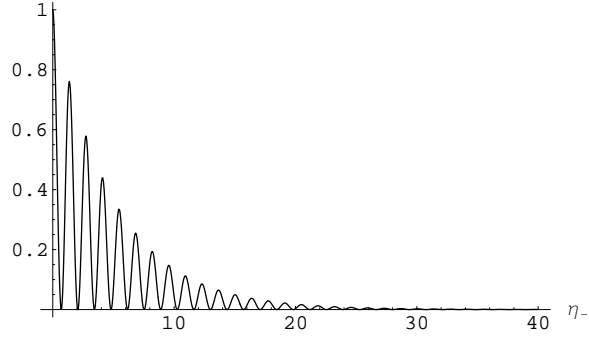}
\caption{The semiclassical part of the two-point funciton,
$\left<\right.Q^2\left.\right>_{\rm a}^{\rm cl} = \bar{Q}^2$
(see Eq.$(\ref{barQ})$ and below). Its behavior is quite different from
the quantum part shown in the previous figures. Here we take $q_0 =1$,
with other parameters unchanged. }
\label{QQCLvseta}
\end{figure}

Note that the above two-point functions with respect to the coherent
state are independent of the proper acceleration $a$.
$\left<\right. Q(\eta)^2\left.\right>_{\rm a}^{\rm qm}$ and the
variance (squared uncertainty) of $Q$,
\begin{equation}
  \left<\right. \Delta Q (\eta)^2\left.\right> \equiv
  \left<\right.  [Q (\eta)-\bar{Q}(\eta)]^2\left.\right>
  =  \left<\right. Q (\eta)^2\left.\right>_{\rm v}+
  \left<\right.Q(\eta)^2\left.\right> _{\rm a}^{\rm qm}
\end{equation} 
have been shown in FIG. \ref{QQvseta}. $\left<\right.
Q(\eta)^2\left.\right>_{\rm a}^{\rm qm}$ decays exponentially due
to the dissipation of the zero-point energy to the field. As
$\left<\right.Q(\eta)^2 \left.\right>_{\rm a}^{\rm qm}$ decays,
$\left<\right.Q(\eta)^2\left.\right> _{\rm v}$ grows and
compensates the decrease, then saturates asymptotically. Similar
behavior can be found in FIG. \ref{PPvseta}, in which 
\begin{equation}
  \left<\right. \Delta \dot{Q}(\eta)^2\left.\right> \equiv
  \left<\right. [\dot{Q}(\eta)-\bar{\dot{Q}}(\eta)]^2\left.\right>
  =  \left<\right. \dot{Q} (\eta)^2\left.\right>_{\rm v}+
  \left<\right.\dot{Q}(\eta)^2\left.\right> _{\rm a}^{\rm qm}
\end{equation} 
is illustrated.
In FIG. \ref{QQCLvseta} we show the semiclassical two-point funciton
$\left<\right.Q^2\left. \right>_{\rm a}^{\rm cl}$. Its behavior is
quite different from the quantum part shown in the previous figures.

\subsection{Late-time variances and the proper acceleration}

\begin{figure}
\includegraphics[width=8cm]{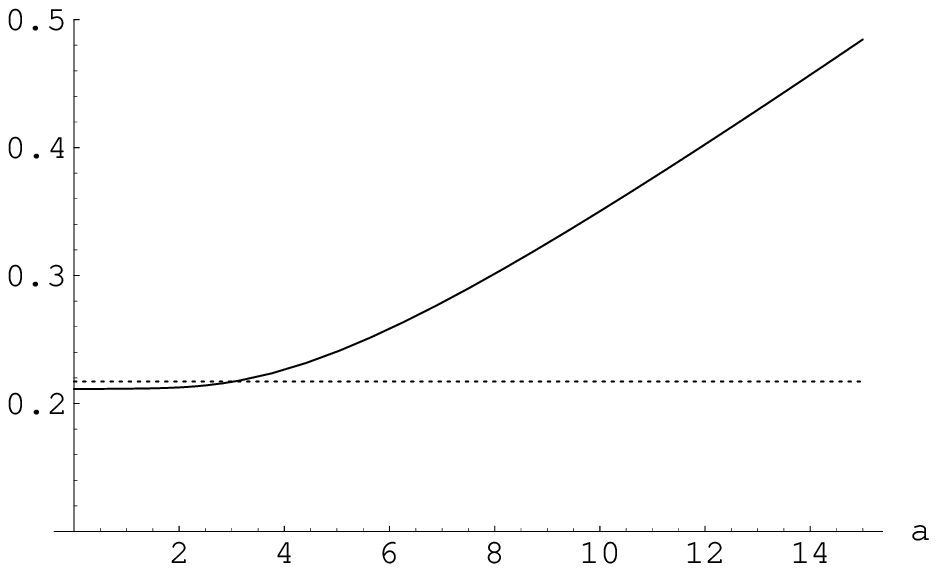}
\caption{$\left<\right. \Delta Q(\infty)^2\left.\right> =
\left<\right. Q(\infty)^2\left.\right>_{\rm v}$ against
the proper acceleration $a$ (solid line, Eq.($\ref{QQsat}$)) with
other parameters the same as FIG. \ref{QQvseta}. For small $a$,
the value of $\left<\right. Q(\infty)^2\left. \right>_{\rm v}$ is
less than $\left<\right. \Delta Q(0)^2\left.\right>=
\left<\right. Q(0)^2\left.\right>_{\rm a}$
(dotted line, see Eq.($\ref{q2bqm})$)). For large $a$,
$\left<\right. Q(\infty)^2\left.\right>_{\rm v}$ is nearly
proportional to $a$.} \label{QQvsa}
\end{figure}

The saturated value $\left<\right. Q(\infty)^2\left.\right>_{\rm v}$ 
in Eq.$(\ref{QQsat})$ is the late-time variance of $Q$, namely, 
$\left<\right. \Delta Q (\infty)^2\left.\right> = \left<\right. 
Q(\infty)^2\left.\right>_{\rm v}$. Its dependence on the proper 
acceleration $a$ is shown in FIG. \ref{QQvsa}.

One can see that, when $a$ is large, $\left<\right.Q(\infty)^2\left.
\right>_{\rm v}$ is nearly proportional to $a$, while in the
zero-acceleration limit $a \to 0$ with $\eta\gg (2\gamma)^{-1}
\ln |\ln a|$, the saturated value goes to a positive number.
From $(\ref{QQsat})$ and $(\ref{q2bqm})$, one finds that
\begin{equation}
  \lim_{a\to 0}{\left<\right. Q(\infty)^2\left.\right>_{\rm v} \over
  \left<\right. Q(0)^2\left.\right>_{\rm a}^{\rm qm}} =
  {i \Omega_r\over \pi\Omega}\ln{\gamma -i\Omega\over \gamma+i \Omega},
\end{equation}
thus $\left<\right.\Delta Q(\infty)^2\left.\right> = 
\left<\right. Q(\infty)^2\left.\right>_{\rm v}$ is smaller than 
$\left<\right. \Delta Q(0)^2\left.\right> = 
\left<\right.Q(0)^2\left.\right>_{\rm a}^{\rm qm}$ for every 
$\gamma >0$ when $a\to 0$. In other words, for a non-accelerated
detector, whose Unruh temperature is zero, the variance of $Q$ in the
detector-field coupled system is still finite and smaller than the one
for the ground state in the free theory.

Actually, $\left<\right. \Delta Q(\infty)^2\left.\right>$ will
become smaller than $\left<\right.\Delta Q(0)^2\left.\right>$ 
whenever $a$ is small enough. Observing FIG. \ref{QQvsa}, 
there is a critical value of $a$ that gives the late-time variance 
identical to the initial one ($a=a_{cr}\approx 3.0447$ in 
FIG. \ref{QQvsa}). Does this mean that the $Q$-component of the 
final wave-packet with $a_{cr}$ is in the original ground state 
of the free theory? The answer is no. What happens is that the 
quantum state of $Q$ has been highly entangled with the quantum 
state of $\Phi$ at late times, and the value of $\sqrt{\left<
\right.\Delta Q(\infty)^2\left.\right>}$ simply represents
the width of the projection of the whole wave-packet (in the
$Q$-$\Phi$ representation of the state) onto the $Q$-axis. There
is actually no factorizable $Q$-component of the wave-packet, and
the final configuration of the wave-packet in $Q$-$\Phi$ space
looks totally different from the initial one. Indeed, with the
same critical value of $a$, $\left<\right. \Delta \dot{Q}
(\infty)^2\left.\right>=\left< \right.\dot{Q}(\infty)^2
\left.\right>_{\rm v}$ is not equal to $\left<\right. \Delta 
\dot{Q}(0)^2\left.\right>=\left<\right. \dot{Q}(0)^2\left.
\right>_{\rm a}^{\rm qm}$ for every $\gamma\not= 0$.

But one can still imagine that, at $\eta=0$, the coherent state
for the free detector is an ensemble of particles with a
distribution function like $|\left<\right.Q\,|\,q \left.\right>|^2$
in $Q$-space. $\sqrt{\left<\right. Q(\eta)^2\left.\right>_{\rm a}^{
\rm qm}}$ is the width of this distribution function. When $\eta>0$,
due to the dissipation which comes with the coupling, all
particles in the ensemble are going to fall into the bottom of
the potential of $Q$, so $\left<\right. Q(\eta)^2\left.
\right>_{\rm a}^{\rm qm}$ shrinks to zero. On the other hand, the
vacuum fluctuations of the field act like a pressure which can push
the ensemble of particles outwards, so that the width of the 
projection of the wave-packet in $Q$-space, $\sqrt{\left<\right.
\Delta Q^2\left.\right>}$, remains finite. A larger $a$ gives 
a higher Unruh temperature, and a higher outward pressure, so
eventually the wave-packet reaches equilibrium with a wider
projection in the potential well of $Q$.

\subsection{Shift of the ground state energy}

A natural definition of the energy of the dressed detector 
(a similar concept is that of a ``dressed atom", see e.g., 
Ref. \cite{CPP, scully}) is 
\begin{equation}
  E(\eta)\equiv {m_0\over 2}\left[\left<\right.\dot{Q}^2(\eta)\left.\right>
  + \Omega_r^2\left<\right.Q^2(\eta)\left.\right>\right], \label{groundE}
\end{equation}
with $\left<\right.Q^2(\eta)\left.\right> = \left<\right.Q^2(\eta)\left.
\right>_{\rm v}+ \left<\right.Q^2(\eta)\left.\right>_{\rm a}$ and $\left<
\right.\dot{Q}^2(\eta)\left.\right> = \left<\right.\dot{Q}^2(\eta)\left.
\right>_{\rm v}+ \left<\right.\dot{Q}^2(\eta)\left.\right>_{\rm a}$
according to $(\ref{splitQQ})$. In FIGs. \ref{QQvseta}-\ref{QQCLvseta},
one can see that $\bar{Q}$, $\left<\right.\dot{Q}^2(\eta)\left.
\right>_a^{\rm qm}$ and $\left<\right.Q^2(\eta)\left.\right>_a^{\rm qm}$
eventually die out. So the late-time energy of the dressed detector is
\begin{eqnarray}
  E(\infty) &=& {m_0\over 2}\left[ \left<\right. \dot{Q}(\infty)^2
  \left.\right>_{\rm v}+ \Omega_r^2\left<\right. Q(\infty)^2
  \left.\right>_{\rm v}\right]\nonumber\\
  &=& {\hbar\over 2\pi}\left\{ a-2{\rm Re}\left[ (\gamma+ i\Omega)
  \psi_{\gamma+i\Omega}\right]-2\gamma \ln a\right\}
\end{eqnarray}
from $(\ref{QQsat})$ and $(\ref{PPsat})$. This is actually the true
ground-state energy of the dressed detector, with the vacuum fluctuations
of the field incorporated.
The first term in $E(\infty)$ could be interpreted as the total
energy of a harmonic oscillator in thermal bath, $k_B T_U$, with the
Unruh temperature $T_U = \hbar a /2\pi k_B$.

The ground-state energy of the dressed detector is not identical to the one
for the free detector, $E_0 = \hbar\Omega_r/2$. In particular, if $a$ is 
small enough, the subtracted $E(\infty)$ is lower than $E_0$, though there 
is an ambiguity of a constant in determining the value of the energy. 
This is analogous to the Lamb shift in atomic physics
\cite{Milonni,scully,MPB93}.

\section{Two-point functions of the quantum field}

Similar to the two-point functions of the detector, for
the initial quantum state $(\ref{initQS})$, the two-point function
of $\Phi$ could be split into two parts,
\begin{eqnarray}
  \left<\right.\hat{\Phi}(x)\hat{\Phi}(x')\left.\right>  &=&
    \left< 0_M\right.|\left< \right. q\left.\right|
    \left[\Phi_a(x)+\Phi_b(x)\right]\left[\Phi_a(x')+\Phi_b(x')\right]
    \left|\right.q\left. \right> |\left. 0_M\right> \nonumber\\ &=&
    G_{\rm v}(x,x') +  G_{\rm a}(x,x'), \label{twopt}
\end{eqnarray}
where, from $(\ref{defPhi})$ and $(\ref{phiab})$,
\begin{eqnarray}
  G_{\rm v}(x,x') &\equiv&\left< \right. q\,|\,q\left.\right>
    \left< 0_M\right.| \Phi_a(x) \Phi_a (x')|\left. 0_M\right>
    =\int {\hbar d^3k\over (2\pi)^3 2\omega}f^{(+)}(x;{\bf k})
    f^{(-)}(x';{\bf k}),\label{GA}\\
  G_{\rm a}(x,x') &\equiv&
    \left< 0_M\right.|\left. 0_M\right>\left< \right. q\left.\right|\Phi_b(x)
    \Phi_b(x')\left|\right.q\left. \right> =
    {\hbar\over 2\Omega_r m_0}f^a(x)f^{a*}(x').   \label{Gbgen}
\end{eqnarray}
Eqs.$(\ref{Phi+})$-$(\ref{fretard})$ and $(\ref{Phibfinal})$
suggest that $G_{\rm v}$ accounts for the back reaction of the vacuum
fluctuations of the scalar field on the field itself, while $G_{\rm a}$
corresponds to the dissipation of the zero-point energy of the internal degree
of freedom of the detector.

Substituting $(\ref{Phi+})$ into $(\ref{GA})$, $G_{\rm v}$ can be decomposed
into four pieces,
\begin{equation}
  G_{\rm v}(x,x') =G_{\rm v}^{00}(x,x')+ G_{\rm v}^{01}(x,x')+
    G_{\rm v}^{10}(x,x')+ G_{\rm v}^{11}(x,x'), \label{Gvij}
\end{equation}
in which $G_{\rm v}^{ij}$ are defined by
\begin{equation}
  G_{\rm v}^{ij}(x,x') \equiv \int {\hbar d^3k\over (2\pi)^3 2\omega}
    f^{(+)}_i(x;{\bf k})f^{(-)}_j(x',{\bf k}),\label{defGAij}
\end{equation}
with $i,j=0,1$. $G_{\rm v}^{00}$ is actually the Green's function
for free fields, which should be subtracted to obtain the renormalized
Green's function for the interacting theory, namely,
\begin{equation}
  G_{\rm ren}(x,x')\equiv  \left< \right. \hat{\Phi}(x)
  \hat{\Phi}(x')\left.\right> - G_{\rm v}^{00}(x,x').
\label{Grentotal}
\end{equation}
Since $G_{\rm v}^{01}(x,x') = [G_{\rm v}^{10}(x',x)]^*$ by
definition, it is sufficient to calculate  $G_{\rm v}^{11}(x,x')$
and $G_{\rm v}^{10}(x,x')$ in the following.

The structure of $G_{\rm v}^{11}$ is quite simple. Comparing 
$(\ref{Phi+1})$, $(\ref{QQvorigin})$ and the definition 
$(\ref{defGAij})$, one concludes that
\begin{equation}
  G_{\rm v}^{11}(x,x') = {\lambda_0^2\over (2\pi)^2 a^2 XX'}
\left<\right. Q(\tau_-)Q(\tau_-')\left.\right>_{\rm v}.\label{G11v}
\end{equation}
The result $(\ref{QtQt'})$ can be substituted directly to get the 
coincidence limit of $G_{\rm v}^{11}$.

By definition, $G_{\rm v}^{10}$ accounts for the interference 
between the retarded solution $f^{(+)}_1$ and the free solution
$f^{(+)}_0$. Since we are interested in the coincidence limit
of $G_{\rm v}^{10}$, and $f_1^{(+)}$ vanishes in the L-wedge 
($U>0$, $V<0$) and P-wedge ($U, V<0$) of Minkowski space, below 
only the $G_{\rm v}^{10} (x,x')$ with $x$ and $x'$ in the 
F-wedge ($U, V>0$) and R-wedge would be calculated.

It has been given in Ref.\cite{lin03b} that
\begin{eqnarray}
  \int{\hbar d^3 k\over (2\pi)^3 2\omega}\varphi(\kappa,{\bf k})
  e^{i\omega t-i{\bf k\cdot x}} &=& \int { e^{i\kappa\tau}d\tau/(2\pi)^3
  \over \left(x^1 - z^1(\tau)\right)^2+\rho^2 -
  \left(t-z^0(\tau)+i\epsilon\right)^2} \nonumber\\
  &=& {i\over 2\pi a X}{\hbar\over (1-e^{-2\pi \kappa/a})}
  \left[e^{i\kappa\tau_-}-Z(\kappa)e^{i\kappa\tau_+}\right],
\end{eqnarray}
where $\epsilon>0$, $Z(\kappa)=1$ and $e^{-\pi\kappa/a}$ for $x$ 
in R and F-wedges
\footnote{Note that there is a transcription oversight in Eq.(54) of
Ref.\cite{lin03b}. An overall factor $-1$ should have been
on the right hand side of $\phi^F$.}, respectively, $X$ and $\tau_-$
were defined in $(\ref{defX})$ and $(\ref{deftaum})$, and
\begin{equation}
  \tau_+ \equiv {1\over a}\ln {a\over 2|U|}
    \left(X-UV+\rho^2+a^{-2}\right).
\end{equation}
Hence, from $(\ref{Phi+1})$ and $(\ref{freephi})$, one has
\begin{equation}
  G_{\rm v}^{10}(x,x') =
  {\hbar\lambda_0^2\theta(\eta_-)\over (2\pi)^3 m_0 a^2 XX'}
  \int {d\kappa\over 1-e^{-2\pi \kappa/a}}\sum_j
  {c_j e^{-i\kappa(\tau_0-\tau'_0)} \over \kappa-i w_j}
  \left[ e^{w_j \eta_-}-e^{-i\kappa\eta_-}\right]
  \left[ e^{i\kappa\eta_-'}-Z(\kappa)e^{i\kappa\eta_+'}\right]
\label{formalG10}
\end{equation}
with $\eta_{\pm}(x)\equiv \tau_\pm(x) -\tau_0$. The
coincidence limit of $G_{\rm v}^{10}$ reads
\begin{eqnarray}
  G_{\rm v}^{10}(x,x) &\equiv& \lim_{x'\to x}{1\over 2}
  \left( G_{\rm v}^{10}(x,x')+ G_{\rm v}^{10}(x',x)\right)\nonumber\\
  &=&{\hbar\lambda_0^2\theta(\eta_-)\over (2\pi)^3m_0 \Omega a^2 X^2}
  {\rm Re} \left\{  i\psi_{\gamma+i\Omega}
  +{ia\over\gamma+i\Omega+a} \left[ e^{-(\gamma+i\Omega+a)\eta_-}
  F_{\gamma+i\Omega}(e^{-a\eta_-}) \right.\right.\nonumber\\
  & & \left.\left.-(\pm) e^{-(\gamma+i\Omega) \eta_- -a\eta_+ }
  F_{\gamma+i\Omega}(\pm e^{-a\eta_+}) \pm e^{-a(\eta_+ -\eta_-)}
  F_{\gamma+i\Omega}(\pm e^{-a(\eta_+ -\eta_-)}) \right]\right\},
\label{G10v}
\end{eqnarray}
with $+$ and $-$ for $x$ in R and F-wedges, respectively.
Near the event horizon $U\to 0$, $\eta_+$ diverges, and the last two terms in
$(\ref{G10v})$ vanish.

As for $G_{\rm a}(x,x')$,
since $f_0^{\rm a}=0$, one has $G_{\rm a}^{01}=G_{\rm a}^{10}=0$,
and only $G_{\rm a}^{11}$ contributes to $G_{\rm a}$. Inserting
$(\ref{Phibfinal})$ into $(\ref{Gbgen})$ and comparing with
$(\ref{QQBgen})$, one finds that
\begin{equation}
  G_{\rm a}(x,x') = {\lambda_0^2 \over (2\pi)^2 a^2 X X'}
    \left<\right.Q(\tau_-)Q(\tau'_-)\left.\right>_{\rm a} . \label{Gball}
\end{equation}
It can also be divided into a quantum part $ G_{\rm a}^{\rm
qm}(x,x')$ and a semiclassical part $G_{\rm a}^{\rm cl}(x,x')$
according to $(\ref{QQBgen})$ and below.

\subsection{Effects due to the interfering term}

In our (3+1) dimensional UD detector theory,
the coincidence limit of the quantum part of $G_{\rm ren}$ reads
\begin{eqnarray}
  G_{\rm ren}^{\rm qm}(x,x) &\equiv& G_{\rm ren}(x,x)-
  G_{\rm a}^{\rm cl}(x,x)\nonumber\\
  &=& G_{\rm a}^{\rm qm}(x,x)+ G_{\rm v}^{11}(x,x)+
    G_{\rm v}^{10}(x,x)+G_{\rm v}^{01}(x,x),  \label{Grenxx}
\end{eqnarray}
owing to $(\ref{twopt})$, $(\ref{Gvij})$ and $(\ref{Grentotal})$.
Collecting the results in $(\ref{G11v})$, $(\ref{G10v})$ and
$(\ref{Gball})$, it is found that $G_{\rm ren}^{\rm qm}(x,x)$ is
singular at $x\to y(\tau)$, and one has to be more cautious.

As can be seen from $(\ref{G11v})$ and $(\ref{Gball})$, $G_{\rm
v}^{11}(x,x)$ and $G_{\rm a}^{\rm qm}(x,x)$ look like the squares of
the retarded field with effective squared scalar charge
$\left<\right.Q(\tau)^2\left.\right>_{\rm v}$ and
$\left<\right.Q(\tau)^2\left.\right>_{\rm a}^{\rm qm}$,
respectively. Since the detector is accelerating, these two terms do
carry radiated energy (this will be shown explicitly later). The
interfering term $G_{\rm v}^{10}(x,x)+ G_{\rm v}^{01}(x,x)$, is more
intriguing: At first glance, it acts like a polarization in the
medium, which screens the radiation field carried by $G_{\rm
v}^{11}(x,x)$ and $G_{\rm a}^{\rm qm}(x,x)$. However, the
interfering term $G_{\rm v}^{10}+G_{\rm v}^{01}$ does not respond to
$G_{\rm a}^{\rm qm}$ at all -- it is independent of $f^a$ and
impervious to any information about  the quantum state of $Q$. Hence
the interfering term cannot be interpreted as the polarization in
the medium. The total effect is simply a destructive interference
between the field induced by the vacuum fluctuations, and the vacuum
fluctuations themselves. For physical interpretations one should
group $G_{\rm v}^{10}+G_{\rm v}^{01}$ and $G_{\rm v}^{11}$ together
and leave $G_{\rm a}^{\rm qm}$ alone.

\begin{figure}
\includegraphics[width=8cm]{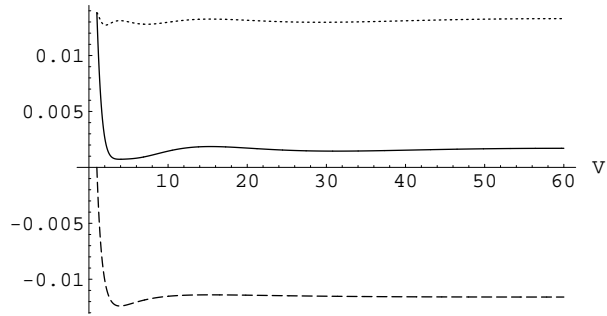}
\caption{Plot of $G_{\rm v}^{10}+G_{\rm v}^{01}$ (dashed
line), $G_{\rm v}^{11}+G_{\rm a}^{\rm qm}$ (dotted line) and
their sum (solid line) against $V$ near the event horizon $U=0$.
Other parameters are the same as those in FIG. \ref{QQvseta}.
One can see the feature that positive $G_{\rm v}^{11}+G_{\rm
a}^{\rm qm}$ is screened by negative $G_{\rm v}^{10}+G_{\rm
v}^{01}$.}
\label{Gtot2}
\end{figure}

\begin{figure}
\includegraphics[width=8cm]{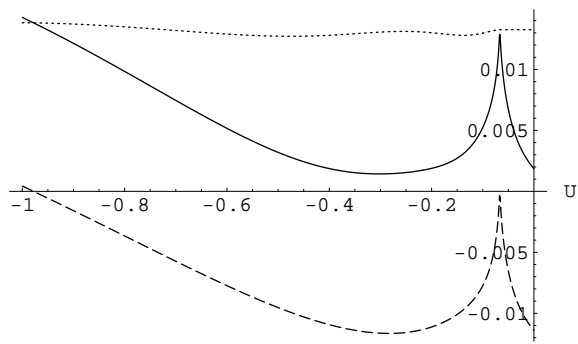}
\caption{Plot of $X^2(G_{\rm v}^{10}+G_{\rm v}^{01})$ (dashed
line), $X^2(G_{\rm v}^{11}+G_{\rm a}^{\rm qm})$ (dotted line) and
their sum (solid line) against $U$ at $V=15$ and $\rho=0$. The
cusp in the right ($U\approx -0.07$ in this plot) locates at the
position of the detector with $X^2(G_{\rm v}^{10}+ G_{\rm
v}^{01})=0$. It is due to the weaker divergence than $1/X^2$ for
$G_{\rm v}^{10}+G_{\rm v}^{01}$.}
\label{Gtot3}
\end{figure}

These quantities, together with their sum, are illustrated in
FIGs. \ref{Gtot2} and \ref{Gtot3}. In FIG. \ref{Gtot2}, one can see
that, soon after the coupling is turned on at $V=1/a$, $G_{\rm
v}^{10}+G_{\rm v}^{01}$ build up and pull the solid curve down.
Observing $(\ref{G10v})$, the time scale (in proper time of the
detector) of this pull-down is about $1/(\gamma+a)$, which is
shorter than the time scale $1/2\gamma$ for
$\left<\right.Q(\tau)^2\left.\right>_{\rm v}$ and
$\left<\right.Q(\tau)^2\left. \right>_{\rm a}^{\rm qm}$, since we
take $\gamma=0.1 < a =1$ here. In FIG. \ref{Gtot3}, one can also see
that $G_{\rm a}^{\rm qm}+G_{\rm v}^{11}$ diverge as $X^{-2}$ around
the trajectory, while the divergence of $G_{\rm v}^{10}+G_{\rm
v}^{01}$ as $X\to 0$ is a bit weaker than $X^{-2}$, such that
$X^2(G_{\rm v}^{10}+G_{\rm v}^{01})$ goes to zero on the trajectory
of the detector.

\subsection{What exactly is the ``vacuum polarization cloud" 
around the detector?}

In prior work for (1+1)D spacetime \cite{RSG} the counterpart of 
$G_{\rm ren}(x,x)$ has been considered as evidence for the existence 
of a ``vacuum polarization cloud" around the detector \cite{Unr92, 
MPB93, RHA}. This is because $G_{\rm ren}(x,x)$ around the detector 
does not vanish even after the system reaches equilibrium, it 
exchanges particles with the detector, and the mean energy it 
carries is zero. Nevertheless, vacuum polarization is a concept 
pertinent to field-field quantum interacting systems.  In quantum
electrodynamics, electrons are described in terms of a field, which
distributes in the whole spacetime, so vacuum polarization is
pictured as the creation and annihilation  of virtual
electron-positron pairs everywhere in spacetime. These virtual
electron-positron pairs do modify the field strength around the
location of a point charge, yielding a non-vanishing variance of the
electromagnetic (EM) field. But in the UD detector theory, at the
level of precision explored here, the detector-field interaction
(hence the virtual processes) only occurs on the trajectory of the
detector. There is no virtual detector or scalar charge at any
spatial point off the location of the UD detector.

Hence in UD detector theory ``vacuum polarization cloud" is not a
precise description of $G_{\rm ren}(x,x)$ in steady state. At
late times $G_{\rm ren}(x,x)$ simply shows the characteristics of
the field in the true vacuum state, in contrast to
$\left<\right.Q(\infty)^2\left.\right>$ for the true ground state of
the detector.


\section{Radiated Power}

\begin{figure}
\includegraphics[width=8cm]{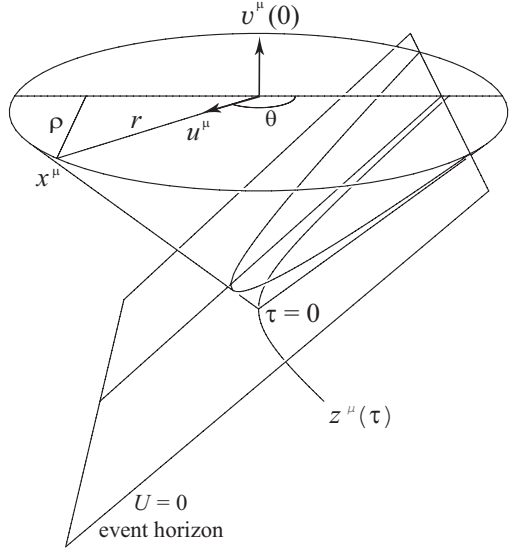}
\caption{Definitions of $\rho$, $\theta$ and $r$.}
\label{defr}
\end{figure}

In classical theory, the modified stress-energy tensor for a
massless scalar field $\Phi$ in Minkowski space is \cite{BD}
\begin{equation}
  T_{\mu\nu}[\Phi(x)] =\left( 1-2\xi\right)\Phi_{,\mu}\Phi_{,\nu}-2\xi\Phi
  \Phi_{;\mu\nu}
  + \left( 2\xi -{1\over 2}\right)g_{\mu\nu}\Phi^{,\rho}\Phi_{,\rho}
  +{\xi\over 2}g_{\mu\nu}\Phi\Box\Phi ,\label{Tmnxi}
\end{equation}
where $\xi$ is a field coupling parameter, set to zero here. Denote
$v^\mu = dz^\mu/d\tau$ as the four velocity of the detector, and
define the null distance $r$ and the spacelike unit vector $u^\mu$
by $x^\mu - z^\mu(\tau_-)\equiv r (u^\mu + v^\mu(\tau_-))$
\cite{rohr} with normalization $u_\mu u^\mu =1$ and $v_\mu u^\mu=0$
(see FIG. \ref{defr}).
Then the stress-energy tensor for the classical retarded field 
$\Phi_{\rm ret}$ induced by the UD detector moving along the 
trajectory $z^\mu(\tau_-)$ can be written as \cite{lin03b}
\begin{eqnarray}
 && T_{\mu\nu}[\Phi_{\rm ret}(x)]_{\xi=0} = {\lambda_0^2\over (4\pi)^2}
   \theta(\eta_-)\left\{{1\over r^4} Q^2(\tau_-)\left( -{1\over 2}
   g_{\mu\nu}+u_\mu u_\nu\right)\right.\nonumber\\&&+{1\over r^3}
   Q(\tau_-)\left[ \dot{Q}(\tau_-) +Q(\tau_-) a_\rho u^\rho\right]
    \left( -g_{\mu\nu}+2 u_\mu u_\nu +u_\mu v_\nu + v_\mu u_\nu\right)
  \nonumber\\ &&+\left.
  {1\over r^2}\left[ \dot{Q}(\tau_-)+ Q(\tau_-)
   a_\rho u^\rho\right]^2(u_\mu +v_\mu)(u_\nu +v_\nu) \right\} .
\label{Tmnret}
\end{eqnarray}
The $O(r^{-2})$ term in the above expression corresponds to the
radiation field, which carries radiated power given by \cite{rohr}
\begin{equation}
  {dW^{\rm rad}\over d\tau_-} = -\lim_{r\to\infty}\int r^2 
    d\Omega_{\rm II} u^\mu T_{\mu\nu}v^\nu (\tau_-)
  ={\lambda_0^2\over (4\pi)^2} \int d\Omega_{\rm II} \left[ 
  \dot{Q}(\tau_-)+Q(\tau_-) a_\rho u^\rho\right]^2 \label{dWdtOII}
\end{equation}
to the null infinity of Minkowski space. Here the $Q$-term
corresponds to dipole radiation ($l=1$, $m=0$ in multipole 
expansion of the radiation field) with the angular distribution 
$a_\rho u^\rho = a\cos \theta$ 
\footnote{The angular distribution of the dipole
radiation emitted by a scalar charge depends on the choice of $\xi$
in the modified stress-energy tensor. For $\xi=1/6$, the angular
distribution is $a_\rho u^\rho = a \sin \theta$, which is the same
as the one for the EM radiation emitted by a electric charge in
electrodynamics.}, while the $\dot{Q}$-term corresponds to monopole 
radiation ($l=0$) isotropic in the rest frame instantaneously 
for the UD detector at $\tau_-$. 
The solid angle $d\Omega_{\rm II}$ could be further integrated out, 
then one obtains the classical radiation formula
\begin{equation}
  {dW^{\rm rad}\over d\tau_-} = {\lambda_0^2\over 4\pi}\left[
  \dot{Q}^2(\tau_-) + {a^2\over 3}Q^2(\tau_-)\right],
\label{Eloss}
\end{equation}
which is the counterpart of the Larmor formula for EM
radiation. The second term is the usual radiation formula for the 
massless scalar field emitted by a constant, point-like scalar 
charge in acceleration \cite{RenWein}.

Naively, one may expect that the quantum version of the radiation
formula could look like $(\lambda_0^2/4\pi)[
\left<\right.\dot{Q}^2(\tau_-)\left.\right> +
(a^2/3)\left<\right.Q^2(\tau_-)\left.\right>]$. In the following, we
shall calculate the quantum expectation value of the flux
$T_{\mu\nu}$, from which we will see that the quantum radiation
formula is more complicated than expected.

\subsection{expectation value of the stress-energy tensor}

The expectation value of the renormalized stress-energy tensor $\left<
T_{\mu\nu}\right>_{\rm ren} $ is obtained by calculating
\begin{equation}
   \left< T_{\mu\nu}[\Phi(x)]\right>_{\rm ren} = \lim_{x'\to x}
   \left[ {\partial\over \partial x^\mu}{\partial\over\partial x'^\nu}
   -{1\over 2}g_{\mu\nu} g^{\rho\sigma}{\partial\over \partial x^\rho}
   {\partial\over\partial x'^\sigma}\right] G_{\rm ren}(x,x'),
\label{expTmn}
\end{equation}
according to $(\ref{Tmnxi})$ with $\xi=0$. With the results in the
previous section, it is straightforward to obtain $\left< T_{\mu\nu}
\right>_{\rm ren}$ induced by the UAD:
\begin{eqnarray}
  \left< T_{\mu\nu}[\Phi(x)]\right>_{\rm ren} &=&
  {\lambda_0^2\theta(\eta_-)\over(2\pi)^2 a^2 X^2}\left[
  g_\mu{}^\rho g_\nu{}^\sigma -{1\over 2}g_{\mu\nu}g^{\rho\sigma}\right]
  \times\nonumber\\ && \left[ \eta_{-,\rho}\eta_{-,\sigma}
  \left<\right. \dot{Q}(\tau_-)^2\left.\right>_{\rm tot} + {X_{,\rho}
  X_{,\sigma}\over X^2}\left<\right.Q(\tau_-)^2\left.\right>_{\rm tot}
  \right.\nonumber\\ && - {X_{,\rho}\over X}\eta_{-,\sigma}\left<\right.
  Q(\tau_-)\dot{Q}(\tau_-)\left.\right>_{\rm tot}-
  \eta_{-,\rho}{X_{,\sigma}\over X}\left< \right.\dot{Q}(\tau_-)
   Q(\tau_-)\left.\right>_{\rm tot}\nonumber\\
  && \left. + \left(\eta_{-,\rho}\eta_{+,\sigma}+
  \eta_{+,\rho}\eta_{-,\sigma}\right){\hbar\Theta_{+-}\over 2\pi m_0} 
  - \left({X_{,\rho}\over X}\eta_{+,\sigma}  + 
  \eta_{+,\rho}{X_{,\sigma}\over X}\right) 
  {\hbar\Theta_{+X}\over 2\pi m_0}\right].\label{Tmnren}
\end{eqnarray}
Upon collecting $(\ref{Phi1Phi1})$ and $(\ref{Xterm})$ as well as
those from $G^{\rm a}$. Here $\left<\right.\dot{Q}^2\left.
\right>_{\rm tot}\equiv \left<\right.\dot{Q}^2\left.\right> + 
(\hbar/2\pi m_0)\Theta_{--}$, $\left<\right.Q^2\left.\right>_{\rm tot} 
\equiv\left<\right.Q^2\left.\right>+(\hbar/2\pi m_0)\Theta_{XX}$
and $\left<\right. Q\dot{Q}\left.\right>_{\rm tot}\equiv
\left<\right. Q\dot{Q}\left.\right>+(\hbar/2\pi m_0)\Theta_{X-}$ 
with $\Theta_{ij}$ defined in $(\ref{Th11})$-$(\ref{Thpm})$.
To see the properties of quantum nature, we define the total 
variances by subtracting the semiclassical part from $\left<\right.
\cdots\left.\right>_{\rm tot}$ as 
\begin{eqnarray}
  \left<\right.\Delta Q^2(\tau_-)\left.\right>_{\rm tot} &\equiv& 
    \left<\right.Q^2(\tau_-)\left.\right>_{\rm tot}-\bar{Q}^2(\tau_-) 
    = \left<\right.\Delta Q^2(\tau_-)\left.\right>+
    {\hbar\Theta_{XX}(\tau_-)\over 2\pi m_0}. \label{Qtot} \\
  \left<\right. \Delta\dot{Q}^2(\tau_-)\left.\right>_{\rm tot} 
    &\equiv& \left<\right.\dot{Q}^2(\tau_-)\left.\right>_{\rm tot}-
    \dot{\bar{Q}}^2(\tau_-) = 
    \left<\right.\Delta\dot{Q}^2(\tau_-)\left.\right>
    +{\hbar\Theta_{--}(\tau_-)\over 2\pi m_0},\label{dotQtot}
\end{eqnarray}
Their evolution against $\eta_-$ are illustrated in FIGs. 
\ref{QQtot1} and \ref{PPtot1}.

In our case, the Minkowski coordinate $(U,V,\rho)$ of a spacetime 
point in F and R-wedge can be transformed to the coordinate 
$(r,\tau_-,\theta)$ by 
\begin{eqnarray}
  \rho &=& r \sin \theta, \label{rhotor}\\
  V &=& r e^{a\tau_-}\left[ 1 + \cos\theta +  (ar)^{-1}\right],
    \label{Vtor}\\
  U &=& r e^{-a\tau_-}\left[ 1 - \cos\theta - (ar)^{-1}\right],
    \label{Utor}
\end{eqnarray}
so that $X = 2r/a$. Also one has
\begin{equation}
  u^\mu (\tau_-) =\left( r\cos\theta\sinh a\tau_-, \,
  r\cos\theta\cosh a\tau_-, \,{x^2\over r}, \,{x^3\over r}\right)
\label{umutau-}
\end{equation}
with $\rho=\sqrt{(x^2)^2+(x^3)^2}$. Now Eq.$(\ref{Tmnren})$ can be 
directly compared with $(\ref{Tmnret})$ and $(\ref{dWdtOII})$. 
One can see clearly that the $\left<\right.\dot{Q}(\tau_-)^2\left.
\right>_{\rm tot}$ term has the same angular distribution as the
one for the $\dot{Q}^2$ term in $(\ref{Tmnret})$, hence would be
recognized as a monopole radiation by the Minkowski observer. 
The angular distributions of the remaining terms in $(\ref{Tmnren})$
are, however, much more complicated because of their dependence on 
$\eta_+(r,\tau_-,\theta)$.

\subsection{Screening}

We have mentioned in the previous section that $G_{\rm a}^{\rm qm}$
and $G_{\rm v}^{11}$ in $(\ref{Gball})$ and $(\ref{G11v})$ carry
radiated energy, now this becomes clear. Observing that what
correspond to $\partial_\mu\partial_{\nu'}[G_{\rm a}^{\rm qm}(x,x')
+G_{\rm v}^{11}(x,x')]$  are those proportional to 
$\left<\right.\cdots\left.\right>$ in
$\left<\right.\cdots\left.\right>_{\rm tot}$ terms of $(\ref{Tmnren})$. 
These terms contribute
a positive flux. Nevertheless, due to the presence of the 
interfering terms $\Theta_{ij}$, most of this positive flux of 
quantum nature will be screened 
when the system reaches steady state as $\eta_-\to\infty$.

\begin{figure}
\includegraphics[width=8cm]{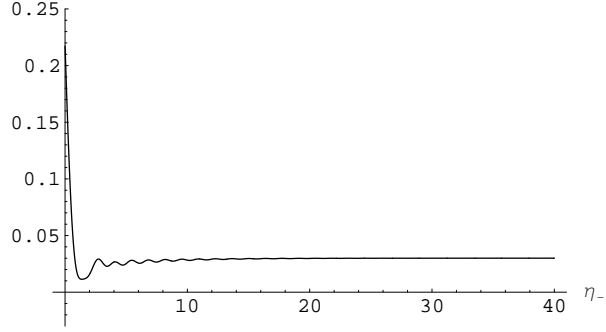}
\caption{The total variance $\left<\right.\Delta Q^2\left. 
\right>_{\rm tot}$ (Eq.$(\ref{Qtot})$) 
near the event horizon for the detector ($\eta_+\to \infty$).
This plot is virtually the same as FIG. \ref{Gtot2} except the 
``time" variable is $\eta_-$ here. The values of parameters are 
still the same as before. The total variance finally saturates
to the value $\hbar a/2\pi m_0 \Omega_r^2$. 
One can compare this plot with FIG. \ref{QQvseta} 
directly and see the suppression.} \label{QQtot1}
\end{figure}

As shown in FIG. \ref{QQtot1}, the total variance $\left<\right.
\Delta Q^2\left.\right>_{\rm tot}$ near the event horizon $U=0$
drops exponentially in proper time (power-law in the Minkowski 
time) after the coupling is turned on. Note that $\left<\right.
\Delta Q^2(\eta_-(x))\left.\right>_{\rm tot}$ is proportional 
to $G_{\rm ren}^{\rm qm}(x,x)$ defined in $(\ref{Grenxx})$, 
and $X$ is independent of $V$ on the event horizon, so 
FIG. \ref{QQtot1} is virtually the same plot as FIG. \ref{Gtot2} 
except that the time variable here is $\eta_-$.
Thus, similar to the behavior of $G_{\rm rem}^{\rm qm}(x,x)$ 
near the event horizon, $\Theta_{XX}$ ($\sim G_{\rm v}^{10}+ 
G_{\rm v} ^{01}$) builds up and the total variance 
$\left<\right.\Delta Q(\tau_-)^2 \left.\right>_{\rm tot}$ is 
pulled down during the time scale $1/(\gamma+a)$ (for $\gamma <a$) 
according to $(\ref{Q^2v})$, $(\ref{q2bqm})$ and $(\ref{Th00})$.
Then $\left<\right.\Delta Q(\tau_-)^2\left.\right>_{\rm tot}$ 
turns into a tail ($\eta_- >1$ in FIG. \ref{QQtot1}) 
which exponentially approaches the saturated value $\hbar a/2\pi m_0
\Omega_r^2$ with the time scale $1/2\gamma$. 

\begin{figure}
\includegraphics[width=8cm]{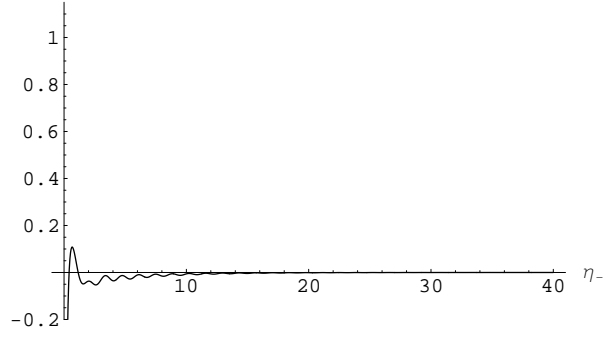}
\caption{The total variance $\left<\right.\Delta \dot{Q}^2
\left.\right>_{\rm tot}$ (Eq.$(\ref{dotQtot})$)
with the same parameters. Note that $\left<\right.\Delta \dot{Q}^2
\left.\right>_{\rm tot}$ is independent of $\eta_+$, and this plot 
is not restricted around the event horizon. One can compare with 
FIG. \ref{PPvseta} and see the suppression.}\label{PPtot1}
\end{figure}

For $\left<\right.\Delta\dot{Q}(\tau_-)^2\left.\right>_{\rm tot}$ 
and $\left<\right.\Delta Q(\tau_-)\Delta \dot{Q}(\tau_-)\left.
\right>_{\rm tot}$, their behaviors are similar (see FIG.
\ref{PPtot1}). In particular,  $\left<\right.\Delta\dot{Q}(\tau_-)^2
\left.\right>_{\rm tot}=\left<\right.\Delta \dot{Q}(\tau_-)^2\left.
\right> + (\hbar /2\pi m_0)\Theta_{--}$ goes to zero at late times 
from $(\ref{dotQ^2v})$, $(\ref{dotQ2aqm})$ and $(\ref{Th11})$ with 
$\gamma\eta_-\gg 1$, so the corresponding monopole radiation 
vanishes after the transient. 

From the calculations of Ref.\cite{lin03b} based on perturbation
theory, it was suggested that the existence of a monopole
radiation could be an experimentally distinguishable evidence of the
Unruh effect. Here we find from a non-perturbative calculation that,
in fact, only the transient of it could be observed. (A comparison
of both results will be given in Sec.VII.) This appears to agree
with the claim that for a UAD in (1+1)D, emitted radiation is only 
associated with nonequilibrium process \cite{CapHJ}. The negative 
tail of $\left<\right.\Delta\dot{Q}(\tau_-)^2\left.\right>_{\rm tot}$ 
in FIG. \ref{PPtot1} and the corresponding quantum  radiation 
could last for a long time with respect to the Minkowski observer 
($\sim V^{-2\gamma/a}$), but this is essentially a transient. 
The interference between the quantum radiation 
induced by the vacuum fluctuations and the vacuum fluctuations 
themselves totally screens the information about the Unruh effect
in this part of the radiation. 

\subsection{Conservation between detector energy and radiation}

What is the physics behind the interfering term in 
$\left<\right.\dot{Q}^2(\eta)\left.\right>_{\rm tot}$?
By inserting our results into $(\ref{groundE})$ and $(\ref{dotQtot})$,
one can show that,
\begin{eqnarray}
  E(\eta_i)-E(\eta_f) &=& {\lambda_0^2\over 4\pi}
    \int_{\eta_i}^{\eta_f} d\eta
    \left<\right.\dot{Q}^2(\eta)\left.\right>_{\rm tot} \nonumber\\
    &=& {\lambda_0^2\over 4\pi} \int_{\eta_i}^{\eta_f} d\eta
    \left[ \dot{\bar{Q}}^2(\eta)+
    \left<\right.\Delta\dot{Q}^2(\eta)\left.\right>
    +{\hbar\Theta_{--}(\eta)\over 2\pi m_0} \right],\label{Econserv}
\end{eqnarray}
for all proper time interval after the interaction is turned on
$(\eta_f > \eta_i >0)$. The left hand side of this equality is the
energy-loss of the dressed detector from $\eta_i$ to $\eta_f$, while
the right hand side is the radiated energy via the monopole radiation 
corresponding to $\left<\right.\dot{Q}^2(\eta)\left.\right>_{\rm tot}$
during the same period. Therefore $(\ref{Econserv})$ is simply a
statement of energy conservation between the detector and the field
{\it in this channel}, and the interfering terms $\Theta_{--}$ must
be included so that $\left<\right.\dot{Q}^2(\eta)\left.\right>_{\rm
tot}$ is present on the right hand side instead of the naively
expected $\left<\right.\dot{Q}^2(\eta)\left.\right>$.
A simpler but more general derivation of this relation is given in
Appendix $\ref{simpleconserv}$. Eq.$(\ref{Econserv})$ also justifies
that $(\ref{groundE})$ is indeed the correct form of the internal 
energy of the dressed detector.

With the relation $(\ref{Econserv})$ we can make two observations
pertaining to results and procedures given before.
First, while the $\Lambda_0$-terms in $\left<\right.Q^2(\tau)\left.
\right>_{\rm v}$ (Eq.($\ref{Q^2v}$)) and $\left<\right.\dot{Q}^2(\tau)
\left.\right>_{\rm v}$ (Eq.($\ref{dotQ^2v}$)) are not included in any
figure of this paper, they are consistent with the conservation law
$(\ref{Econserv})$. Actually the $\Lambda_0$-term in $(\ref{Q^2v})$
satisfy the driving-force-free equation of motion $(\ref{noforceeom})$,
just like the semiclassical $\bar{Q}$ does.

Second, Eq.$(\ref{Econserv})$ implies that all the internal energy
of the dressed detector dissipates via a monopole radiation, and
the external agent which drives the detector along the trajectory
$(\ref{cltraj})$ has no additional influence on this channel.

\subsection{Quantum radiation formula}

\begin{figure}
\includegraphics[width=6cm]{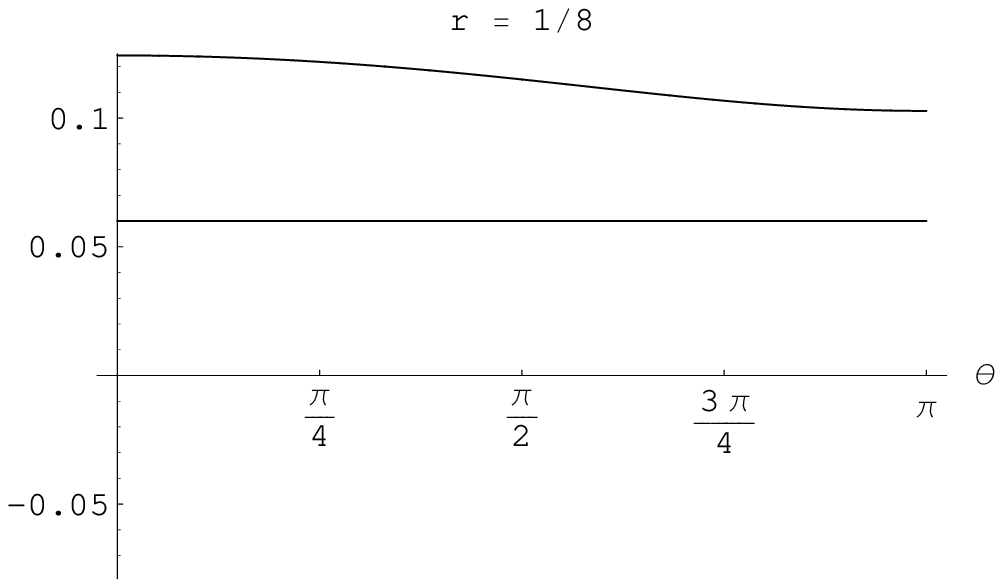}
\includegraphics[width=6cm]{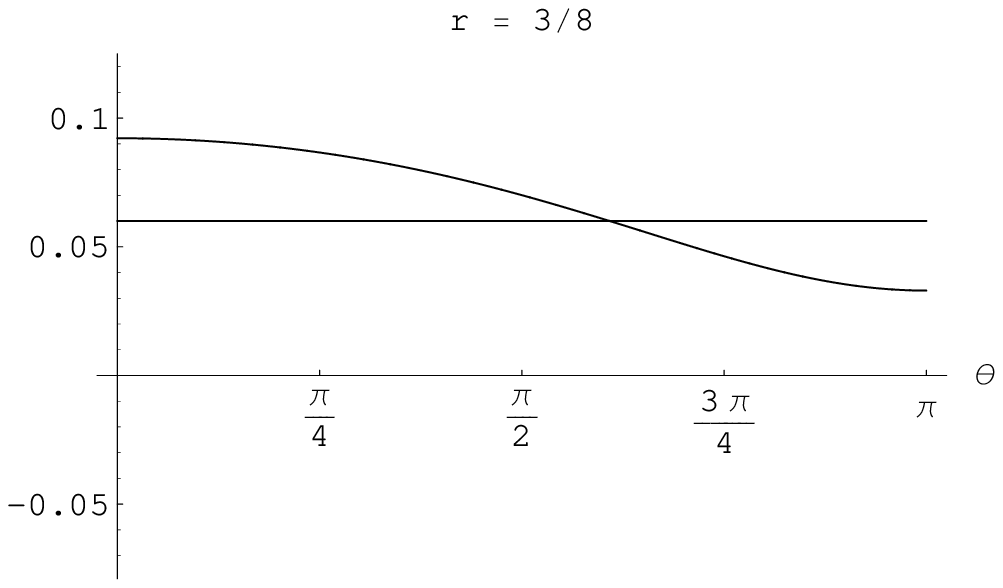}
\includegraphics[width=6cm]{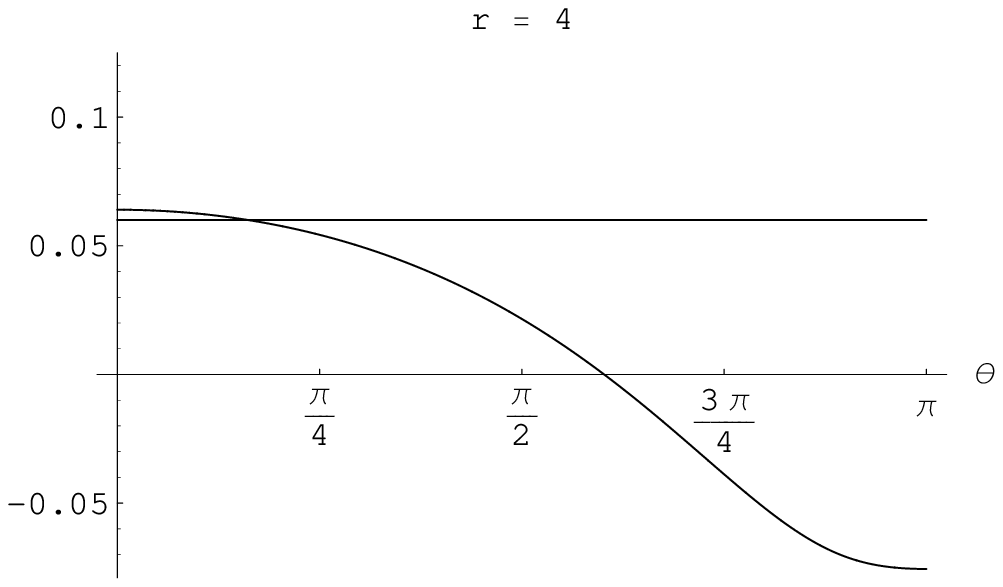}
\includegraphics[width=6cm]{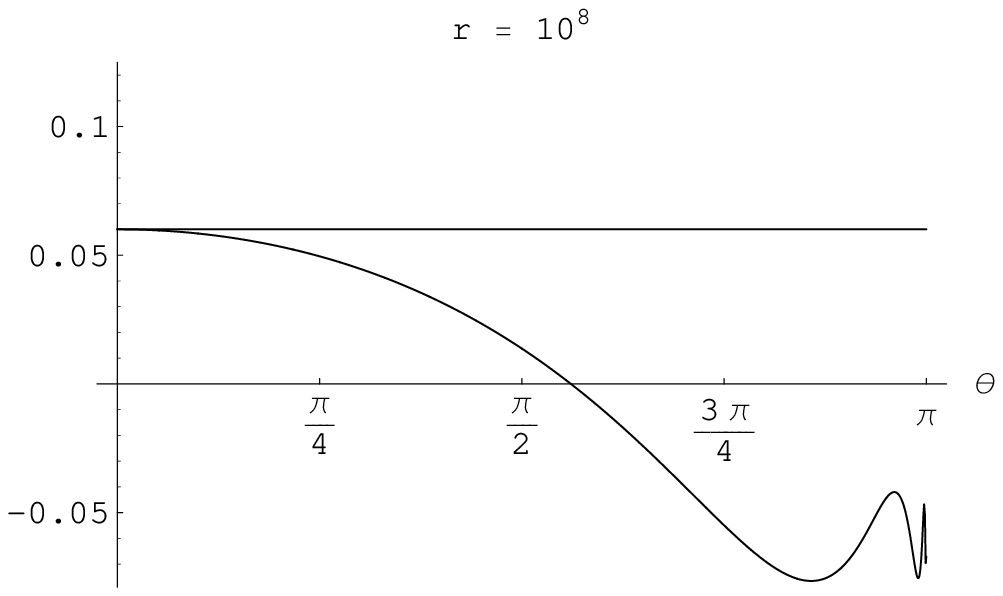}
\caption{Angular distributions of $\left<\right.\Delta Q^2\left.
\right>_{\rm tot}$ at late times (Eq.(\ref{Qtot}) with 
$\gamma\tau_- \gg 1$) as the null distance $r$ increases. 
The values of parameters are the same as before except that here 
we choose $a=2$. 
The horizontal line indicates the saturated value of $\hbar a/
2\pi m_0 \Omega_r^2$ shown in FIG. \ref{QQtot1} . 
When $r$ starts with $0$, $\left<\right. Q^2\left.
\right>_{\rm tot}$ decreases from $\left<\right. Q^2(\infty)\left.
\right>_{\rm v}$ (top-left). The light cone hits the
event horizon at $\theta=\pi$ when $r=1/(2a)=1/4$, and
$\left<\right. Q^2\left.\right>_{\rm tot}$ always has the value
$\hbar a/2\pi m_0 \Omega_r^2$ at the event horizon (top-right, the
curve and the horizontal line intersect right at the event horizon).
As $r$ further increases, $\left<\right. Q^2\left. \right>_{\rm
tot}$ sinks more and more (bottom-left), and some oscillations begin
to develop near $\theta=\pi$. Finally $\left<\right.
Q^2\left.\right>_{\rm tot}$ is non-vanishing at the null infinity
$r\to\infty$, with the value smaller than $\hbar a/2\pi m_0
\Omega_r^2$ whenever $\theta\not= 0$ (bottom-right).} 
\label{ThXX1}
\end{figure}

\begin{figure}
\includegraphics[width=6cm]{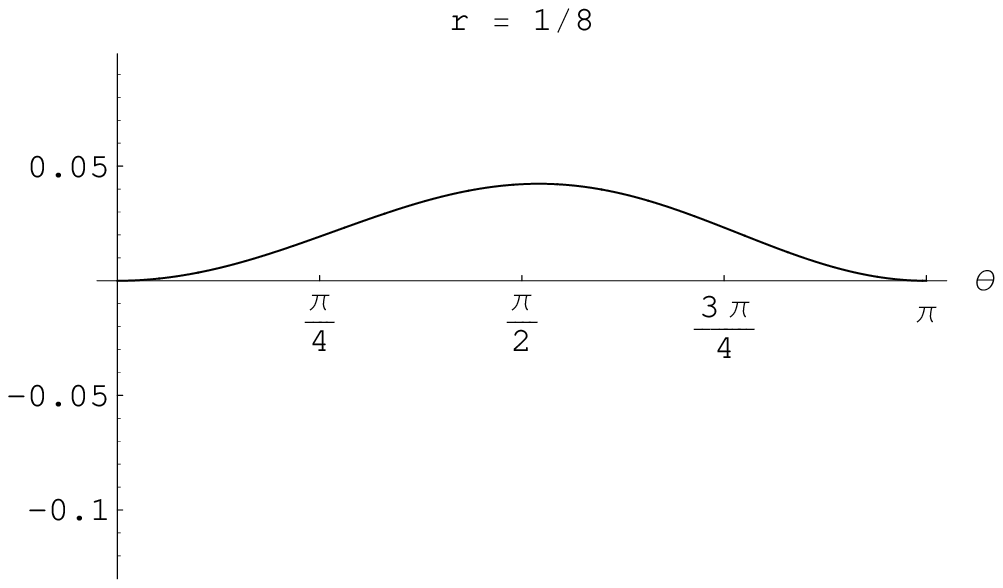}
\includegraphics[width=6cm]{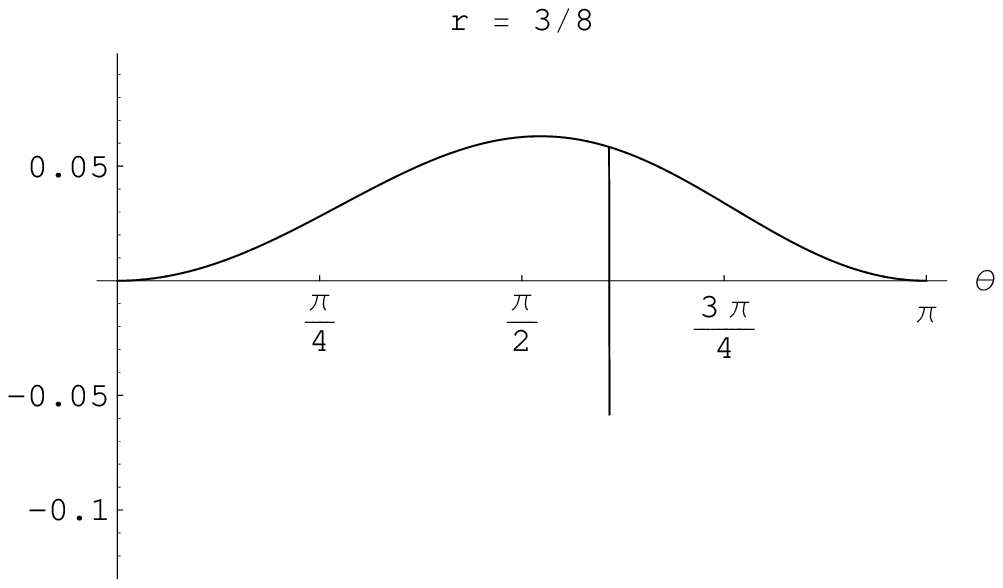}
\includegraphics[width=6cm]{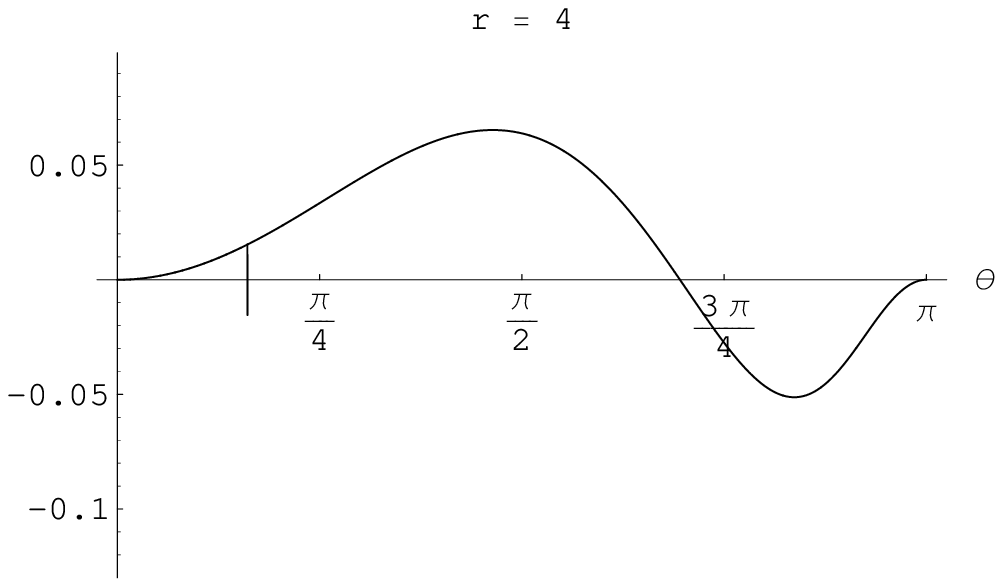}
\includegraphics[width=6cm]{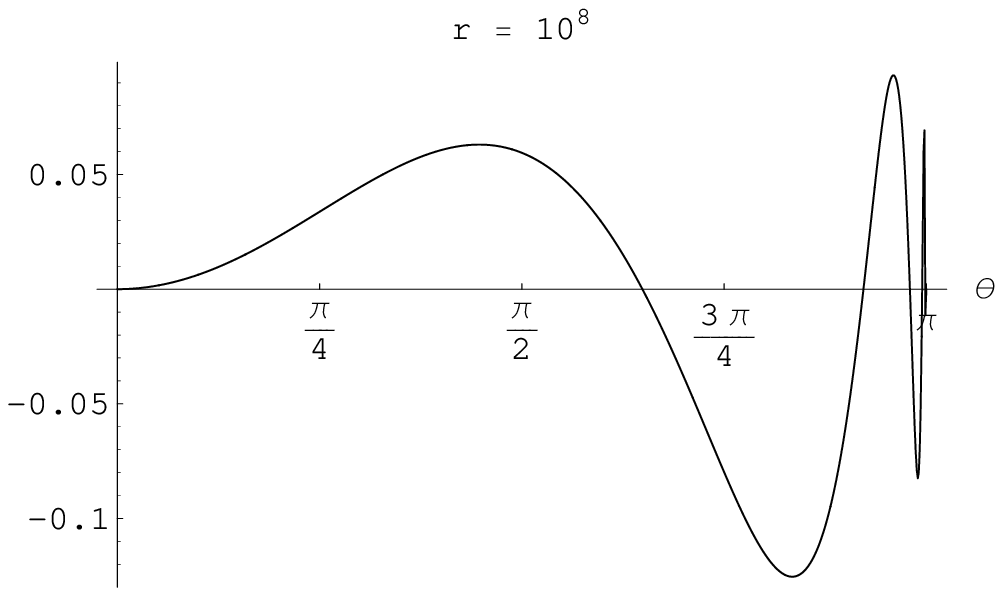}
\caption{Angular distributions of $\chi \equiv u^\mu(
\eta_-^\mu\eta_+^\nu +\eta_+^\mu\eta_-^\nu)v^\nu\Theta_{+-}$ 
(Eq.($\ref{Thpm}$)) at late times as $r$ increases. Parameters 
are the same as those in FIG. \ref{ThXX1}. When $r$
starts with $0$, $\chi$ grows from $0$ (top-left).
After the light cone hits the event horizon, $\chi$ keeps regular 
on the event horizon (top-right, the vertical line indicates 
the position of the event horizon). As $r$ further increases, a 
nodal point enters from the right (bottom-left). 
Then more and more nodal points can be seen. Finally $\chi$ is 
non-vanishing at the null infinity $r\to \infty$, with
infinitely many nodal points close to $\theta=\pi$ (bottom-right).} 
\label{Thpm1}
\end{figure}

Transforming $(\ref{Tmnren})$ to the form of $(\ref{Tmnret})$ by 
applying $(\ref{rhotor})$-$(\ref{Utor})$, one can calculate the 
radiation power
\begin{equation}
  \left<{dW^{\rm rad}\over d\tau_-}\right> = -\lim_{r\to\infty}
    \int r^2 d\Omega_{\rm II} \,\,u^\mu
    \left<\right.T_{\mu\nu}\left.\right>_{\rm ren} v^\nu (\tau_-)
\label{Qradformula0}
\end{equation}
following a similar argument in classical theory. Before calculating,
let us observe the behavior of the steady-state $r^2 u^\mu 
\left<\right.T_{\mu\nu}\left.\right>_{\rm ren} v^\nu$ in the forward
light cone. As $r$ increses, the developments of two terms in 
late-time $r^2 u^\mu \left<\right.T_{\mu\nu}\left.\right>_{\rm ren} 
v^\nu$ are illustrated in FIGs. \ref{ThXX1} and \ref{Thpm1}. 
It turns out that both are regular and non-vanishing 
at the null infinity of Minkowski space ($r\to \infty$) even in 
steady state ($\gamma\eta_- \gg 1$). FIGs. \ref{ThXX1} and 
\ref{Thpm1} also indicate that, near the the null 
infinity of Minkowski space, almost all the equal-$r$ surface lies 
in the F-wedge, except the region around $\theta=0$ is still in 
the R-wedge. The contribution to the integral around $\theta=0$ 
can be totally neglected because the value of $r^2 u^\mu 
\left<\right.T_{\mu\nu}\left.\right>_{\rm ren} v^\nu$ is regular 
there while the measure for this portion in the angular integral 
is zero when $r\to \infty$. 
So the radiation power can be written as
\begin{equation}
  \left<{dW^{\rm rad}\over d\tau_-}\right> =
  {\lambda_0^2\over 8\pi} \int_0^\pi d\theta \sin \theta 
    \left\{\left<\right. \dot{Q}^2\left.\right>_{\rm tot} 
    -{\hbar\Theta_{+-}\over \pi m_0}+
   a^2 \cos^2\theta \left<\right. Q^2\left.\right>_{\rm tot} 
   + a\cos\theta \left[\left<\right. \{Q,\dot{Q}\}\left.\right>_{\rm tot}
   - {\hbar\Theta_{+X}\over \pi m_0}\right] \right\},
\label{Qradformula1}
\end{equation}
by inserting $(\ref{Tmnren})$, $(\ref{umutau-})$ and $v^\mu(\tau_-)=
(\cosh a\tau_-,\sinh a\tau_-,0,0)$ into $(\ref{Qradformula0})$.
This is the quantum radiation formula for the massless scalar field
emitted by the UAD in (3+1)D spacetime.

At late times, while $\left<\right.\dot{Q}^2\left.\right>_{\rm tot}$ 
ceases, it still remains a positive radiated power flow
\begin{eqnarray}
  &&\left<{dW^{\rm rad}\over d\tau_-}\right> 
  \stackrel{\gamma\tau_-\to\infty}{\longrightarrow}\nonumber\\
&&  {\hbar\lambda_0^2\over 8\pi^2 m_0 }\int_0^\pi d\theta 
    \sin\theta \left\{ {a\over 2\Omega_r^2}a^2\cos^2\theta
     -a \sin^2{\theta\over 2}
  - {a\over\Omega}\tan^2{\theta\over 2}\,\,{\rm Re}
  \left[{i(\gamma+i\Omega-a\cos\theta)^2\over \gamma+i\Omega +a}
    \,\, F_{\gamma+i\Omega}\left( -\tan^2{\theta\over 2}\right)
    \right] \right\}\nonumber\\ &&= {\hbar\lambda_0^2
    \over 8\pi^2 m_0 }\left\{ {a^3\over 3\Omega_r^2} -a 
  -{2\over 3}\left[ {a^3\over \Omega_r^2}-a  + 2\gamma
    + {\rm Re}\left[{i(\gamma+i\Omega)\over a\Omega}
      \left[ (\gamma+i\Omega)^2-a^2\right] \psi^{(1)}
    \left({\gamma+i\Omega \over a}\right)\right]\right] \right\} 
\label{QMrad}
\end{eqnarray}
to the null infinity of Minkowski space. Thus we conclude that 
there exists a steady, positive radiated power of quantum nature 
emitted by the detector even when the detector is in steady state.

For large $a$, the first term in $(\ref{QMrad})$ dominates, 
and the radiated power is approximately
\begin{equation}
  \left<{dW^{\rm rad}\over d\tau_-}\right> \approx
  {\lambda_0^2 \over 4\pi } {a^2\over 3}{\hbar a\over 2\pi m_0\Omega_r^2}
  \propto a^2 T_U,
\end{equation}
where $T_U$ is the Unruh temperature. This could be interpreted as 
a hint of the Unruh effect. Note that it does not originate from 
the energy flux that the detector experiences in Unruh effect, 
since the internal energy of the dressed detector is conserved 
only in relation to the radiated energy of a monopole radiation 
corresponding to $\left<\right.\dot{Q}^2\left.\right>_{\rm tot}$. 
Learning from the EM radiation
emitted by a uniformly accelerated charge \cite{rohr, Boulware}, we
expect that the above non-vanishing radiated energy of quantum
origin could be supplied by the external agent we introduced in the
beginning to drive the motion of the detector.
Further analysis on
the quantum radiations of the detector involving the dynamics of the
trajectory is still on-going.

\section{Comparison with Earlier Results}

We can recover the results in Ref.\cite{lin03b}, which is obtained by
perturbation theory, as follows. In Ref.\cite{lin03b}, the first order
approximation of the flux $\left<\right. T^{tU}\left.\right>_{\rm ren}$ 
through the event horizon $U=0$ has been calculated. Here, the 
expectation value of $T^{tU}$ near the event horizon reads, from 
$(\ref{Tmnren})$,
\begin{eqnarray}
  && \left<\right. T^{tU}\left.\right>_{\rm ren}|_{U\to 0}\equiv
  \lim_{x'\to x}\left[ 2\partial_V\partial_{V'}+  {1\over 2}
  \partial_\rho\partial_{\rho'}\right] G_{\rm ren}(x,x')|_{U\to 0}\nonumber\\
   &=& {2\lambda_0^2\over (2\pi)^2 a^4 \left(\rho^2+a^{-2}\right)^2}
\theta\left[-{1\over a}\ln {a\over V}\left(\rho^2+a^{-2}\right)-\tau_0\right]
    \left\{ {1\over V^2}\left<\right.\dot{Q}(\tau_-)^2\left.\right>_{\rm tot}+
    \right.\nonumber\\ && \left.\left. {\rho^2\over (\rho^2+a^{-2})^2} \left[
    a^2 \left<\right.Q(\tau_-)^2\left.\right>_{\rm tot} + a \left< \right.
    \dot{Q}(\tau_-)Q(\tau_-)+Q(\tau_-)\dot{Q}(\tau_-)\left.\right>_{\rm tot}
    + \left<\right.\dot{Q}(\tau_-)^2\left.\right>_{\rm tot}\right] \right\}
    \right|_{U\to 0},\label{powerflux}
\end{eqnarray}
in which $\Theta_{+-}$ and $\Theta_{+X}$ terms vanish.
By letting $\gamma \to 0$ with $\eta_-$ finite, then taking 
$\eta_-\to \infty$, the total variance 
$(\ref{dotQtot})$ becomes
\begin{eqnarray}
  \left<\right. \Delta\dot{Q}(\tau_-)^2\left.\right>_{\rm tot}
  &\stackrel{\gamma\to 0}{\longrightarrow}&
    {\hbar\Omega_r\over 2 m_0} -{\hbar\over 2\pi m_0}\left\{
    a + {2a \cos\Omega_r\eta_-\over e^{a\eta_-}-1} -2\Omega_r
    {\rm Re}\left[  {ae^{(-a+i\Omega_r)\eta_-}\over \Omega_r +ia}
    F_{i\Omega_r}(e^{-a\eta_-})- i \psi_{-i\Omega_r}\right]\right\}
    \nonumber\\ &\stackrel{\eta_-\to \infty}{\longrightarrow}&
    {\hbar\Omega_r\over m_0(1-e^{2\pi\Omega_r/a})}, \label{wrong}
\end{eqnarray}
owing to $(\ref{dotQ^2v})$, $(\ref{dotQ2aqm})$ and $(\ref{Th11})$.
This is identical to the corresponding part of Eq.(66) in
Ref.\cite{lin03b},
\begin{equation}
  2\sum_E |\left< E_0\right.|Q(0)|\left.E\right>|^2 {\varepsilon^2
  \over 1-e^{2\pi\varepsilon}}, \label{monoradGRL}
\end{equation}
by noting that there, $m_0=1$, $\sum_E|\left< E_0\right.|Q(0)
|\left.E\right>|^2 = \left< E_0\right.| Q(0)^2 |\left.E_0\right>
=\hbar/2\varepsilon$, and $\varepsilon$ there is equal to
$\Omega_r/a$ here. 

The monopole radiation corresponding to $(\ref{monoradGRL})$ looks
like a constant negative flux since $\varepsilon >0$. 
Accordingly it was concluded in Ref.\cite{lin03b} that 
such a quantum monopole radiation could be
experimentally distinguishable from the bremsstrahlung of the
detector. At first glance this constant monopole radiation seems
to contradict the knowledge gained from (1+1)D results. But actually
similar results for (1+1)D cases were also obtained by Massar and
Parentani (MP) \cite{MP}, who found that a detector initially
prepared in the ground state and coupled to a field under a smooth
switching function does emit radiation during thermalization.
They pointed out that the radiated flux in what they refered to as
the ``golden rule limit" ($\eta\to\infty$ with $\gamma\eta$ small,
while the switching function becomes nearly constant) is
approximately a constant negative flux for all $V>0$ \footnote{A
word of caution in terminology: Note that the golden rule yields
Markovian dynamics which is the prevailing case under equilibrium
conditions, as during uniform acceleration. To avoid confusion in
its connotation, it is perhaps simpler and more precise just to
state the condition explicitly. }. In spite of the long interaction
time and the nearly constant radiated flux, the detector will remain
in dis-equilibrium.

Note that the initial conditions in \cite{lin03b} are similar to
those in Sec.II of MP, and the limiting condition for obtaining
$(\ref{wrong})$ is exactly what MP assumed there.
Hence, the constant negative flux in \cite{lin03b} is essentially a
transient effect, which exists only in the period that the above
stated condition holds. When the interaction time $\eta$ exceeds
$O(\gamma^{-1})$, this approximation breaks down.
To obtain the correct late-time behavior, one should take the limit
$\eta_-\to \infty$ before $\gamma \to 0$.
Then $\left<\right. \Delta \dot{Q}(\tau_-)^2\left.\right>_{\rm tot}$ 
goes to zero.

\section{Summary}
In this paper, we consider the Unruh-DeWitt detector theory in (3+1)
dimensional spacetime. A uniformly accelerated detector is modeled
by a harmonic oscillator $Q$ linearly coupled with a massless scalar
field $\Phi$. The cases with the coupling constant $|\lambda_0|$ less
than the renormalized natural frequency $|\Omega_r|$ of the detector
are considered.

We solved exactly the evolution equations for the combined system of
a moving detector coupled to a quantum field in the Heisenberg
picture, and from the evolution of the operators we can obtain
complete information on the combined system. For the case that the
initial state is a direct product of a coherent state for the
detector and the Minkowski vacuum for the field, we worked out the
exact two-point functions of the detector and similar functions of
the field.  By applying the coherent state for the detector, we can
distinguish the classical behaviors from others. The quantum part of
the coincidence limit of two-point functions, namely, the variances
of $Q$ and $\Phi$, are determined by the detector and the field 
together.

From the exact solutions, we were able to study the complete process
from the initial transient to the final steady state. In particular,
we can identify the time scales of transient behaviors analytically.
When the coupling is turned on, the zero-point fluctuations of the free
detector dissipates exponentially, then the vacuum fluctuations take
over. The time scales for both processes are the same. Eventually
the variance of $Q$ saturates at a finite value, where the
dissipation of the detector is balanced by the input from the
vacuum fluctuations of the field. Even in the zero-acceleration
limit, the variances of $Q$ and $\dot{Q}$, thus the ground state
energy of the interacting detector, shifted from the ones for the
free detector. This fluctuations-induced effect share a similar
origin with that of the Lamb shift.

The variance of $Q$ yields an effective squared scalar charge, which
induce a positive variance in the scalar field. This variance of the
field contributes a positive radiated energy at the quantum
level. However, the interference between the vacuum fluctuations and
the retarded solution induced by the vacuum fluctuations screens 
part of the emission of quantum radiation. The time scale of the
screening is proportional to $1/(\gamma+a)$ for $a>\gamma$, where
$a$ is the proper acceleration and $\gamma$ is the damping constant
proportional to $\lambda^2$. After the screening the
renormalized Green's functions of the field are still non-zero in
steady state.

A quantum radiation formula determined at the null infinity of
Minkowski space has been derived. We found that even in steady state
there exists a positive radiated power of quantum nature emitted 
by the uniformly accelerated UD detector.
For large $a$ the radiated power is proportional to $a^2 T_U$, where 
$T_U$ is the Unruh temperature.
This could be interpreted as a hint of the Unruh effect.
However, the nearly constant negative flux obtained in
Ref.\cite{lin03b} for (3+1)D case is essentially a transient effect.

Only part of the radiation is connected to the internal energy of 
the detector. The total energy of the dressed detector and 
the radiated energy of a monopole radiation from the detector is
conserved for every proper time interval 
after the coupling is turned on. The external agent which drives the
detector's motion has no additional influence on this channel. 
Since the corresponding monopole radiation of quantum nature ceases 
in steady state, the hint of the Unruh effect in the 
late-time radiated power is not directly from the energy flux 
experienced by the detector in Unruh effect. 
This extends the result in Ref.\cite{CapHR, RSG} that there is
no emitted radiation of quantum origin in Unruh effect in (1+1)
dimensional spacetime.


Since all the relevant quantum and statistical information about the
detector (atom) and the field can be obtained from the results
presented here, when appropriately generalized, they are expected to
be useful for addressing issues in atomic and optical schemes of
quantum information processing, such as quantum decoherence,
entanglement and teleportation. These investigations are in
progress.

$\,$

\noindent{\bf Acknowledgments} BLH thanks Alpan Raval and Mei-Ling
Tseng for discussions in an earlier preliminary attempt on the 3+1
problem. We appreciate the referees' queries and suggestions for
improvements on the presentation of this paper. This work is
supported in part by the NSC Taiwan under grant NSC93-2112-M-001-014
and by the US NSF under grant PHY03-00710 and PHY-0426696.

\begin{appendix}

\section{Derivatives of Two-Point Functions of the Field}

From $G_{\rm v}^{11}$ in $(\ref{G11v})$, it is easy
to see that
\begin{eqnarray}
  & &\partial_\mu\partial_{\nu'}G_{\rm v}^{11}(x,x') =
  \int{\hbar d^3 k\over (2\pi)^3 2\omega}\partial_\mu f_1^{(+)}(x)
  \partial_{\nu'}f_1^{(-)}(x') \nonumber\\
  &=&  {\lambda_0^2\over (2\pi)^2 a^2 XX'}\theta(\eta_-)\theta(\eta_-')
  \left[ {X_{,\mu}X'_{,\nu}\over X X'}
  \left<\right. Q(\tau_-)Q(\tau_-')\left.\right>_{\rm v}+
  \eta_{-,\mu}\eta'_{-,\nu}\left<\right. \dot{Q}(\tau_-)
  \dot{Q}(\tau_-')\left.\right>_{\rm v}\right. \nonumber\\ & & \left.
  - {X_{,\mu}\over X}\eta'_{-,\nu} \left<\right. Q(\tau_-)
  \dot{Q}(\tau_-')\left.\right>_{\rm v} -\eta_{-,\mu}{X'_{,\nu}\over X'}
   \left<\right. \dot{Q}(\tau_-)Q(\tau_-')\left.\right>_{\rm v} \right].
\label{Phi1Phi1}
\end{eqnarray}
Note that the $\delta$-functions at $\eta_- =0$, coming from the
derivative of the step functions, have been neglected here.

With $G_{\rm v}^{10}+G_{\rm v}^{01}$, one can write down in closed
form of the interfering terms in the R-wedge of the Rindler space.
Under the coincidence limit, it looks like,
\begin{eqnarray}
  &&\lim_{x'\to x}{1\over 2}\left\{\partial_\mu \partial_{\nu '}
  \left[G_{\rm v}^{10}(x,x')+G_{\rm v}^{01}(x,x')\right] +
  (x\leftrightarrow x')\right\}\nonumber\\&=&
  {\hbar\lambda_0^2\theta(\eta_-)\over(2\pi)^3 m_0 a^2 X^2}
  \left[\eta_{-,\mu}\eta_{-,\nu}\Theta_{--}+
  {X_{,\mu}X_{,\nu}\over X^2}\Theta_{XX}
  -{X_{,\mu}\over X}\eta_{-,\nu}\Theta_{X-} - \eta_{-,\mu}{X_{,\nu}\over X}
  \Theta_{-X}\right.\nonumber\\ & & \left. + \eta_{-,\mu}\eta_{+,\nu}\Theta_{-+}+
  \eta_{+,\mu}\eta_{-,\nu}\Theta_{+-} - {X_{,\mu}\over X}\eta_{+,\nu}
  \Theta_{X+} - \eta_{+,\mu}{X_{,\nu}\over X}\Theta_{+X}\right].\label{Xterm}
\end{eqnarray}
where
\begin{eqnarray}
\Theta_{--} &=& -
  4\gamma(\Lambda_1-\ln a) -a +
  {2 a\over\Omega}{e^{-\gamma\eta_-}\over e^{a\eta_-}-1}
  \left(\gamma\sin\Omega\eta_- -\Omega\cos\Omega\eta_-\right)\nonumber\\
& & -{2\over \Omega} {\rm Re} \left[i(\gamma+i \Omega)^2\psi_{\gamma+i\Omega}
  + {ia(\gamma+i \Omega)^2\over\gamma+i\Omega+a}e^{-(\gamma+i\Omega+a)\eta_-}
  F_{\gamma+i\Omega}(e^{-a\eta_-})\right],\label{Th11}\\
\Theta_{XX} &=& 
  {2\over \Omega} {\rm Re} \left[
  i\psi_{\gamma+i\Omega}+ {ia\over\gamma+i\Omega+a}
  e^{-(\gamma+i\Omega+a)\eta_-} F_{\gamma+i\Omega}(e^{-a\eta_-})
  \right]+2 {\cal F}_0(x),\label{Th00}\\
\Theta_{-X}&=& \Theta_{X-} = -{a\over\Omega}{e^{-\gamma\eta_-}\over
  e^{a\eta_-}-1} \sin\Omega\eta_- + {\cal F}_1(x),\label{ThmX}\\
\Theta_{+X} &=& \Theta_{X+} = {a\over\Omega}{e^{-\gamma\eta_-}\over
  \pm e^{a\eta_+}-1} \sin\Omega\eta_- - {\cal F}_1(x),\label{ThpX}\\
\Theta_{+-} &=& \Theta_{-+}= -{a\over\Omega}{e^{-\gamma\eta_-}\over
  \pm e^{a\eta_+}-1} \left(\gamma\sin\Omega\eta_-
  -\Omega\cos\Omega\eta_-\right)- {a \over \pm e^{a(\eta_+-\eta_-)}-1}
  +{\cal F}_2(x),\label{Thpm}
\end{eqnarray}
with
\begin{equation}
  {\cal F}_n(x) \equiv \pm {a\over \Omega} {\rm Re} \left\{
  {i(\gamma+i\Omega)^n\over\gamma+i\Omega+a}\left[ 
  e^{-a\eta_+ -(\gamma+i\Omega)\eta_- } F_{\gamma+i\Omega}(\pm e^{-a\eta_+})
  -e^{-a(\eta_+-\eta_-)}F_{\gamma+i\Omega}(\pm e^{-a(\eta_+-\eta_-)})
  \right] \right\},
\end{equation}
with $+e^{a\eta_+}$ and $-e^{a\eta_+}$ for $x$ in R and F-wedge, respectively.
Note that $\Theta_{--}$ is independent of $\eta_+$. $\Theta_{XX}$ is actually
proportional to $G_{\rm v}^{10}$ in $(\ref{G10v})$. Another observation is
that, combining $(\ref{Phi1Phi1})$ and $(\ref{Xterm})$, one finds that the
divergent $\Lambda_1$ term in $\Theta_{--}$ is canceled by the one in
$\left<\right.\dot{Q}^2(\eta_-)\left.\right>_{\rm v}$.

As for $G^{\rm a}$, the result is similar to $(\ref{Phi1Phi1})$
with $\left<\cdots\right>_{\rm v}$ being replaced by
$\left<\cdots\right>_{\rm a}$. Again it splits into the quantum and
semiclassical parts as in Sec.IV B.

\section{A simpler derivation of the conservation law}
\label{simpleconserv}

The conservation law $(\ref{Econserv})$ is directly obtained by
arranging our somewhat complicated results. It can also be
derived in a simpler way as follows.

From the definition of the ground-state energy $(\ref{groundE})$
of the dressed detector, its first derivative of $\tau$ is
\begin{equation}
  \dot{E}(\tau) = m_0{\rm Re}\left[
  \left<\right.\dot{Q}(\tau)\ddot{Q}(\tau)\left.\right> +\Omega_r^2
  \left<\right.Q(\tau)\dot{Q}(\tau)\left.\right>\right]
\end{equation}
Introducing the equations of motion $(\ref{eomq2})$ and
$(\ref{eomqb})$ to eliminate $\ddot{Q}(\tau)$, one has
\begin{eqnarray}
 \dot{E}(\tau) &\sim& m_0 {\rm Re} \left\{ \left<\dot{Q}(\tau)
  \left[-2\gamma\dot{Q}(\tau)-\Omega_r^2 Q(\tau) +
  {\lambda_0\over m_0}\Phi_0(y(\tau))\right]\right>_{\rm v} +
  \Omega_r^2 \left<\right.Q(\tau)\dot{Q}(\tau)\left.\right>_{\rm v}\right.
  \nonumber\\& & \left.+ \left<\dot{Q}(\tau)\left[-2\gamma\dot{Q}(\tau)-
  \Omega_r^2 Q(\tau)\right]\right>_{\rm a}+ \Omega_r^2
\left<\right.Q(\tau)\dot{Q}(\tau)\left.\right>_{\rm a}\right\}\nonumber\\
  &=& -2\gamma m_0 \left<\right.\dot{Q}^2(\tau)\left.\right> + \lambda_0
  {\rm Re}\left<\right.\dot{Q}(\tau)\Phi_0(y(\tau))\left.\right>_{\rm v},
\label{dotenergy}
\end{eqnarray}
where the last term is a short-hand for
\begin{equation}
  \lambda_0 {\rm Re} \int {\hbar d^3 k\over (2\pi)^3 2\omega}
  \dot{q}^{(+)}(\tau;{\bf k})f^{(-)}_0(z(\tau);{\bf k}).\label{lastterm}
\end{equation}
(There is an additional term $\lambda_0 \dot{\bar{Q}}(\tau)\Phi_{\rm in}
(y(\tau))$ if a background field $\Phi_{\rm in}(x) =
\left<\right.\Phi_0(x)\left. \right>_{\rm v}$ is present.)
Substituting $(\ref{q+1})$ and $(\ref{f0+zt})$ into $(\ref{lastterm})$,
integrating out $d^3 k$ with the help of Eq.$(\ref{trick1})$,
then comparing with $\partial_\mu\partial'_\nu G^{10}(x,x')$
in $(\ref{Xterm})$ where $G^{10}$ has the formal
expression $(\ref{formalG10})$, one finds that $(\ref{lastterm})$
is actually identical to $-\hbar\gamma \Theta_{--}(\tau)/\pi$.
Hence
\begin{equation}
  \dot{E}(\tau) = -2\gamma m_0 \left<\right.\dot{Q}^2(\tau)
  \left.\right>_{\rm tot}.
\end{equation}
Integrating both side from $\tau=\tau_i$ to $\tau_f$, one ends up
with Eq.$(\ref{Econserv})$.

\end{appendix}


\begin{references}

\bibitem{CTDG}
C. Cohen-Tannoudji, J. Dupont-Roc, and G. Grynberg, {\it
Atom-Photon Interactions: Basic Processes and Applications}
(Wiley, New York, 1992).

\bibitem{CTmovatom}
C. Cohen-Tannoudji, {\it Atomic Motion in Laser Light} eds. J.
Dalibard, J. M. Ramond, and J. Zinn-Justin {\bf Les Houches,
Session LIII, 1990} (Elsevier, Amsterdam, 1991).

\bibitem{Milonni}
P. W. Milonni, {\it The Quantum Vacuum: An Introduction to Quantum
Electrodynamics} (Academic Press, Boston, 1994).

\bibitem{CPP}
G. Compagno, R. Passante and F. Persico, {\it Atom-Field
Interactions and Dressed Atoms} (Cambridge University Press, Cambridge,
1995).

\bibitem{scully} See, for example, M. O. Scully and M. S. Zubairy,
{\it Quantum Optics}, (Cambridge University Press, Cambridge, 1997).

\bibitem{BD} N. D. Birrell and P. C. W. Davies, {\it Quantum Fields in
Curved Space} (Cambridge University Press, Cambridge, 1982),
Chapter 2.

\bibitem{DeW75}B. S. DeWitt, Phys. Rep. {\bf 19C}, 297 (1975).

\bibitem{BH}
J. D. Bekenstein, Phys. Rev. D7, 2333 (1973).

\bibitem {Haw75}
S.W. Hawking, Commum. Math. Phys. {\bf 43}, 199 (1975).

\bibitem{Jackson}
J. D. Jackson, {\it Classical Electrodynamics} (Wiley, New York,
1983).

\bibitem{Boulware} D. G. Boulware, Ann. Phys. (N.Y.) {\bf 124},
169 (1980).

\bibitem{Unr76} W. G. Unruh, Phys. Rev. {\bf D14}, 870 (1976).

\bibitem{DavFul}
S.A. Fulling, P.C.W. Davies, Proc. R. Soc. {\bf A348},
393 (1976). P. C. W. Davies and  S. A. Fulling, Proc. R. Soc.
{\bf A356}, 237 (1977).


\bibitem{Chen}
\textit{Quantum Aspects of Beam Physics}, 18th Advanced IFCA Beam
Dynamics Workshop, edited by Pisin Chen (World Scientific,
Singapore, 2001).

\bibitem{ChenTaj} P. Chen and T. Tajima, Phys. Rev. Lett.
{\bf 83}, 256 (1999).

\bibitem{BelLen}
J. M.  Leinaas, ``Unruh Effect in Storage Rings" in \cite{Chen}

\bibitem{Scullyetal} M. O. Scully, V. V. Kocharovsky, A. Belyanin,
E. Fry and F. Capasso, Phys. Rev. Lett. {\bf 91}, 243004 (2003).

\bibitem{UnrWal}
W. G. Unruh and R. M. Wald,  Phys. Rev. {\bf D29}, 1047 (1984).

\bibitem{Grove}
P. G. Grove, Class. Quan. Grav. {\bf 3}, 801 (1986).

\bibitem{RSG}
D. J. Raine, D. W. Sciama, and P. G. Grove, Proc. R. Soc. 
 {\bf A435}, 205 (1991).

\bibitem{Unr92}
W. G. Unruh, Phys. Rev. {\bf D46}, 3271 (1992).


\bibitem{MPB93} S. Massar, R. Parentani and R. Brout,
Class. Quantum Grav. {\bf 10}, 385 (1993).

\bibitem{AccDetRev}
S. Takagi, Prog. Theor. Phys. Suppl. {\bf 88}, 1 (1986);
V. L. Ginzburg and V. P. Frolov, Sov. Phys. Usp. {\bf 30}, 1073 (1988).

\bibitem{Hin} F. Hinterleitner, Ann. Phys. {\bf 226}, 165 (1993).

\bibitem{AMH}
J. Audretsch and R. M\"uller, Phys. Rev. {\bf D49}, 4056 (1994);
J. Audretsch, R. M\"uller and M. Holzmann, Phys. Lett. {\bf 199A}, 151
(1995).

\bibitem{MP}
S. Massar and R. Parentani, Phys. Rev. {\bf D54}, 7426, 7444  (1996).

\bibitem{CapHR}B. L. Hu and A. Raval, ``Is there Radiation in
Unruh Effect?", in \cite{Chen}[quant-ph/0012135].

\bibitem{RavalPhD}
A. Raval, Ph. D. Thesis, University of Maryland, College Park,
1996 (unpublished).

\bibitem{RHA}
A. Raval, B. L. Hu and J. Anglin, Phys. Rev. {\bf D53}, 7003
(1996).

\bibitem{RHK}
A. Raval, B. L. Hu and D. Koks, Phys. Rev. {\bf D55}, 4795
(1997).

\bibitem{lin03b} S.-Y. Lin, Phys. Rev. {\bf D68}, 104019 (2003).

\bibitem{DeW79} B. S. DeWitt, in {\it General Relativity: an Einstein
Centenary Survey}, edited by S. W. Hawking and W. Israel (Cambridge
University Press, Cambridge, 1979).

\bibitem {CapHJ}B. L. Hu and P. R. Johnson, ``Beyond Unruh
effect: Nonequilibrium quantum dynamics of moving charges" in
\cite{Chen} [quant-ph/0012132].

\bibitem{HRthermal} B. L. Hu and A. Raval, Mod. Phys. Lett. {\bf A11},
2625 (1996).

\bibitem{Parker76} L. Parker, Nature {\bf 261}, 20 (1976)

\bibitem{HuRouraPRL} 
B. L. Hu and A. Roura, Phys. Rev. Lett. 93, 129301 (2004).

\bibitem{JH1}
P. R. Johnson and B. L. Hu, Phys. Rev. {\bf D65}, 065015 (2002).

\bibitem{IARD} P. R. Johnson and B. L. Hu, ``Uniformly Accelerated Charge
in a Quantum Field: From Radiation Reaction to Unruh Effect"
[gr-qc/0501029].

\bibitem{GH1} C. R. Galley and B. L. Hu, Phys. Rev. {\bf D72}, 084023 
(2005); C. R. Galley, B. L. Hu and S.-Y. Lin, ``Electromagnetic and 
gravitational self-force on a relativistic particle from quantum 
fields in curved space" [gr-qc/0603099].

\bibitem{HPZ}
B. L. Hu, J. P. Paz and Y. Zhang, Phys. Rev. {\bf D45}, 2843 (1992).

\bibitem{rohr} F. Rohrlich, {\it Classical Charged Particles}
(Addison-Wesley, Redwood, 1965), Chapter 5.

\bibitem{RenWein} H. Ren and E. J. Weinberg, Phys. Rev. {\bf D49}, 6526
(1993).


\end{references}
\end{document}